\documentclass{article}

\usepackage{PRIMEarxiv}

\usepackage[utf8]{inputenc} 
\usepackage[T1]{fontenc}    
\usepackage{hyperref}       
\usepackage{url}            
\usepackage{booktabs}       
\usepackage{amsfonts}       
\usepackage{nicefrac}       
\usepackage{microtype}      
\usepackage{lipsum}
\usepackage{fancyhdr}       
\usepackage{graphicx}       
\graphicspath{{media/}}     
\usepackage{xcolor}
\usepackage{amsmath}
\usepackage{amsfonts}

\usepackage{amsmath,tabu,booktabs}
\usepackage{stfloats}
\usepackage{caption}
\usepackage{subcaption}
\title{Refraction FWI of a circular shot OBN acquisition in the Brazilian pre-salt region
\thanks{Submitted as a Journal Paper to IEEE Transactions on Geoscience and Remote Sensing} 
}

\author{
  Sérgio Luiz E. F. da Silva$^{1,2,3}$, Felipe T. Costa$^{1}$, Ammir Karsou$^{1}$, Adriano de Souza$^{1}$,  \\ \textbf{Felipe Capuzzo}$^{1}$, \textbf{Roger M. Moreira}$^{1}$, \textbf{Jorge Lopez}$^{3}$, \&  \textbf{Marco Cetale}$^{1}$ \\ \\
  $^{1}$  Grupo de Imageamento Sísmico e Inversão Sísmica (GISIS), Fluminense Federal University, Niter\'oi, RJ, Brazil \\
  $^{2}$ Dipartimento di Scienza Applicata e Tecnologia (DISAT), Politecnico di Torino, Turin, Italy\\
  $^{3}$Institute for Complex Systems of the National Research Council  c/o  Politecnico di Torino, Turin, TO, Italy \\
  $^{4}$ Shell Brasil Petr\'oleo Ltda, Rio de Janeiro, RJ, Brazil \\
  Corresponding author email: \texttt{sergio.dasilva@polito.it} 
}

\begin{document}
\maketitle

\begin{abstract}
We develop a workflow based on full-waveform inversion (FWI) to estimate P-wave velocities in a deepwater Brazilian pre-salt field using the recently introduced circular shot ocean bottom node (OBN) acquisition geometry. Such a geometry comprises a source vessel sailing in large radius concentric circular trajectories and seismic signals are recorded by OBN arrays. The circular shot OBN survey provides mostly refracted waves separately from reflected waves, so the FWI process is mainly driven by diving waves. We introduce a new FWI workflow to analyze non-preprocessed OBN refraction data, which includes automated steps such as data selection solving an Eikonal equation, estimation of a source signature that accounts for ghost and bubble effects, and gradient preconditioning using a non-stationary filter and seismic illumination. We consider two objective functions based on the $L^1$ and $L^2$ norms. The FWI results demonstrated that using our proposed workflow with the $L^1$ norm objective function and the circular OBN survey can lead to an improvement in pre-salt velocity models. Furthermore, using these improved models we construct reverse-time migration (RTM) images of the conventional OBN dataset, showing significant improvements in the salt stratification, the base of salt, and the lateral resolution of the pre-salt area. The Brazilian pre-salt case study demonstrated that the circular shot OBN acquisition maximizes the illumination of deep reservoirs through the ultra-long offset and full-azimuth coverage that prioritizes the recording of diving waves.
\end{abstract}


\section{Introduction}

Ultra-deepwater oil and gas reservoir imaging in complex geology remains a significant challenge due to the poor seismic illumination of these target areas and the high computational efforts for dealing with large physical models. In this context, modern wide-aperture acquisition geometries are essential to register the diving waves from ultra-deep reservoirs, which is vital for robust geophysical inversion and imaging~\cite{Kazei_et_al_2013_FWI}. Guaranteeing low-frequency content and ultra-long offsets in a seismic acquisition is crucial for successfully updating the deep targets in the earth model~\cite{Morgan_et_al_2013_GJI_GeometryAcquisition_FWI}. Ocean bottom node  (OBN) acquisitions naturally emerge as a great option, providing a wide range of azimuths and offsets, recording penetrating diving waves~\cite{Plessix_et_al_2010_FirstBreak_FWI_OBN}.

Interest in the deepwater Brazilian pre-salt has increased due to its giant oil and gas reserves. The deployment of OBN acquisitions has been critical in the characterization and monitoring of pre-salt target areas, improving pre-salt images~\cite{Jouno_et_al_SEG_FWI_OBN_PRESALTBRAZIL_2019}. The circular shot OBN acquisition geometry~\cite{Lopez_et_al_2020_SEG_OBNcircular} is an alternative to maximize the seismic illumination of pre-salt reservoirs using refracted wave energy~\cite{Costa_et_al_2020_SEG_wavepaths}. Such an acquisition geometry comprises a source vessel sailing in concentric circular trajectories while firing into the OBN array. 

Several studies have addressed different aspects of circular geometry using synthetic and real data. Considering synthetic data, Ref. \cite{Costa_et_al_2020_SEG_wavepaths} confirmed through wavepath analysis that the circular geometry is capable of illuminating the ultra-deep reservoirs of the Brazilian pre-salt region. Ref. \cite{daSilva_et_al_SEG_2021_qFWI_BrazilianPreSalt}  verified through synthetic FWI experiments that such wave paths are fundamental in the reconstruction of P-wave velocity models, demonstrating that circular geometry is very important for the seismic inversion of deep targets. Considering the circular-shot OBN field data, Ref. \cite{daSilva_et_al_2022_EAGE_KleinGordon}  explored the effects of spatially varying density on acoustic modeling using a dispersive Klein-Gordon operator. They found that the Klein-Gordon acoustic wave equation allows the calculation of more accurate amplitudes and travel times due to abruptly varying density effects, particularly for free air-water surface-related multiples. Ref. \cite{duarte_et_al_2023_Geophysics} proposed an alternative and complementary workflow to show how well FWI works when using circular geometry to estimate P-wave models, taking into account all the information from the seismograms and an acoustic wave equation of the Klein-Gordon type. They demonstrated the potential of circular geometry in the recovery of structures associated with Brazilian pre-salt reservoirs.
Ref. \cite{costa_et_al_2023_SEG_GdM_Circular} proposed that employing circular geometry for reservoir characterization, along with wavefront reconstruction, automatic data selection using the Eikonal equation, and prior subsurface knowledge, can enhance the imaging of target areas. More recently, Ref. \cite{daSilva_et_al_2024_GeophysProsp_CircularOBNparametrization} demonstrated that properly parameterizing the FWI process through the squared slowness is capable of recovering several layers existing in ultra-deep regions of the Brazilian pre-salt at a reduced computational cost without losing the quality of the retrieved model.

The above-mentioned studies shown that careful planning of OBN acquisition in combination with full-waveform inversion (FWI) can improve the velocity model-building process. FWI is a powerful seismic inversion technique that benefits from the high-quality low-frequency content, full azimuthal coverage and ultra-long offsets provided by OBN datasets~\cite{Martinez_et_al_SEG_FWI_OBN_PRESALTBRAZIL_2020}. FWI is a wave-equation-based seismic imaging technique that aims to generate quantitative subsurface models based on a fully automatic analysis of recorded waveforms (observed data) from a seismic survey~\cite{Virieux_et_al_Geophysics_FWI_introduction_2017}. From a practical point of view, FWI is a nonlinear inverse problem in which the forward problem consists of modeling the propagation of seismic waves by solving a wave equation~\cite{Fichtner_2010_FWI_book}. The inverse problem consists of determining a subsurface model that honors the geology by matching modeled waveforms with the observed waveforms recorded in a seismic survey employing an objective function~\cite{vanLeeuwen_Herrmann_2013_GJI_WRI,Tejero_et_al_2015_GJI_Comparative_ObjectiveFunctions,Warner_Guasch_2016_Geophysics_AWI_Theory,daSilva_et_al_2020_PhysRevE_101_kappaFWI,Pladys_et_al_2021_Geophysics_fiveObjectivesFunctionBrossier,Carvalho_et_al_2021_GJI_nonparametricFWI}. A wide variety of case studies have shown that FWI is a valuable imaging technique for reconstructing subsurface models~\cite{Warner_et_al_2013_Geophysics_AnisotropicFWI,Gorszczyk_et_al_2017_FWI_Nankai,Haacke_et_al_2029_EAGE_MUltiPArFWI,Kamath_et_al_2021_Geophysics_FWI_Multiparameter,daSilva_et_al_2024_GeophysProsp_CircularOBNparametrization}.

Acoustic FWI employing low-frequency data is useful for analyzing waveforms because it requires less computational effort to simulate wave propagation in acoustic media than in elastic media~\cite{Barnes_Charara_2009_Geophysics_AppAcousticFWIMarineData}. For example, Ref.  ~\cite{Sirgue_et_al_2010_FWI_VALHALL} demonstrated that low-frequency FWI in the isotropic acoustic approximation produces impressive velocity models, revealing several geological features, such as a detailed image of gas-charged sediments in the North Sea of the Valhall oil field. References  \cite{Schiemenz_Heiner_2013_GJI_195,Liu_et_al_2013_LeadingEdge_32} also presented successful applications of time-domain isotropic acoustic FWI, showing the feasibility of the acoustic approximation to deal with  industrial-scale real 3D marine data. Indeed, acoustic FWI has been employed for reconstructing subsurface models from pressure data collected by hydrophones (see, for instance, Refs.~\cite{Fu_et_al_2020_GeophysicalProspecting_acousticFWI_OBC_3D,Vigh_et_al_2021_ImpactGeometryFWI,Brenders_et_al_2022_LeadingEdge}).

In this work, we investigate the potential of  3D time-domain acoustic FWI on a  circular shot OBN refraction field dataset. In particular, we consider an OBN dataset from a Brazilian pre-salt field to discuss our proposed workflow to deal with non-preprocessed OBN refraction data, a subset of nodes, and using a single graphics processing unit (GPU) as a computational resource. Also, we present the success and challenges of the target-oriented seismic imaging of the Brazilian ultra-deepwater region by inverting refracted waves using the low-frequency acoustic FWI process and the circular shot OBN acquisition~\cite{Lopez_et_al_2020_SEG_OBNcircular,duarte_et_al_2023_Geophysics}. It is worth noting that Ref. \cite{duarte_et_al_2023_Geophysics} recently demonstrated that this dataset responds well to FWI analyses by considering all waveforms, from primaries to free-surface-related multiples. In this work, we demonstrate how (primary) refracted waves aid FWI processing in the reconstruction of ultra-deep geological structures in the pre-salt region, increasing the horizontal resolution of seismic images, which is of great interest to the industry.

We organize this work as follows. In the next section, we briefly review the main ingredient of acoustic wave propagation and the FWI requirements. In Section \ref{sec:BrazilianpresaltinitialFWIsetup}, we present the Brazilian pre-salt OBN dataset and draw our proposed FWI workflow, which ranges from the refraction data automatic selection using the Eikonal approximation to the FWI objective function optimization issue using adequate gradient preconditioning. In Section \ref{sec:results}, we present the application of our proposed FWI workflow on the circular shot OBN field data from the Brazilian pre-salt region. Finally, we devote the Section \ref{sec:finalremarks} to the final remarks and future implications of our proposal for deep target-oriented FWI.

\newpage
\section{Full-waveform inversion}\label{sec:fwi}

In this section we present the key elements of the acoustic FWI by recalling its theoretical foundations through the objective functions based on $L^1$ and $L^2$ norms. We consider the two most popular and simplest ways to implement objective functions existing in the literature, which are based on the $L^1$ and $L^2$ norms. We consider these criteria to demonstrate that the circular shot OBN acquisition can provide a rich recording of diving waves, and, therefore, that the FWI technique is capable of estimating good subsurface models without relying on sophisticated objective functions.

The main goal of the FWI technique is to obtain an informative subsurface model by comparing modeled and observed waveforms while optimizing an objective function \cite{Virieux_Operto_Geophysics_FWI_overview_2009}. Commonly, metrics based on $L^1$ and $L^2$ norms are employed in geophysical data inverse problems. Such criteria are particular cases of the $L^p$ norm, which for our problem is defined as:
\begin{equation}
    \underset{m}{min} \hspace{.1cm} \phi_{L^p} (m) :=
    \frac{1}{p} \sum_{s,r} \int_{0}^{T}
    \Big|\Big|\Gamma_{s,r} \psi_s(m,t) - d_{s,r}(t)\Big|\Big|^p_p dt,
\label{eq:FWI_lp_norm}
\end{equation}
where $\phi_{L^p}$ represents the $L^p$-objective function, $|| \cdot ||_p$ with $p \geq 1$ is the $L^p$ norm, $m$ denotes the subsurface model parameters, $t \in [0,T]$ represents the time, being $T$ the recording time, $\Gamma_{s,r} \psi_s$ and $d_{s,r}$ are, respectively, the modeled and observed data linked to the seismic source $s$ and the receiver $r$. Cases $p = 1$ and $p = 2$ refer to $L^1$ and $L^2$ norms, the criteria considered in the present work. Although we have not included regularization terms in the objective function \cite{Asnaashari_et_al_2013_Geophysics,Linan_et_al_2023_GJI}, we have carefully preconditioned the gradient of the objective function to avoid additional issues and to ensure numerical stability during the inversion process, as we will explain later.

A crucial part of FWI algorithms is obtaining the modeled waveforms \cite{Virieux_Operto_Geophysics_FWI_overview_2009}. Given a subsurface physical model $m$, acquisition parameters, and a source signature, the modeled data are computed through the forward modeling by solving a wave equation \cite{Fichtner_2010_FWI_book}. In this work we consider the 3D time-domain acoustic modeling using the second-order isotropic acoustic wave equation.  Thus, the modeled wavefield is the pressure field $\psi$ that satisfies the following equation:
\begin{equation}
\nabla^2 \psi_s(\textbf{x},t) 
- m (\textbf{x}) \frac{\partial^2 \psi_s (\textbf{x},t)}{\partial  t^2 }   
= f_s(t) \delta(\textbf{x}-\textbf{x}_s),
\label{eq:wave_equaqtion_time}
\end{equation}
where $\textbf{x} \in \mathbb{R}^3$ denotes the spatial coordinates, $\nabla^2$ is the Laplacian operator, and $f_s(t) \delta(\textbf{x}-\textbf{x}_s)$ represents the seismic source $s$ at the position $\textbf{x} = \textbf{x}_s$ with $\delta$ representing the Dirac Delta function. In this context, the model parameters are the squared slowness for a medium with P-wave velocity $v_p$: $m=1/v_p^2$, in which the spatial coordinate $\textbf{x}$ is implicit in the latter relation. Notice that Eq.~\eqref{eq:wave_equaqtion_time} provides the wavefield $\psi$ everywhere within the physical domain of the subsurface model. However, in practice, the seismic wavefield is only recorded at the receiver positions for each seismic source. Therefore, the modeled data are given by $\Gamma_{s,r} \psi_s$, where $\Gamma_{s,r}$ is a sampling operator onto the receiver $r$ of the source $s$. 

Due to computational efforts, FWI is usually formulated as a gradient-based minimization task \cite{Virieux_et_al_Geophysics_FWI_introduction_2017}. The inversion algorithm iteratively updates the earth model from an initial subsurface model $m_0$ until the distance between modeled and observed data is sufficiently small, according to \cite{Fichtner_2010_FWI_book}
\begin{equation}
    m_{i+1}=m_{i}-\alpha_{i} \, h(m_i) \quad \text{s.t.} \quad \phi(m_{i+1}) < \phi(m_{i}), 
    \label{eq:quasi_Newton_method}
\end{equation}
for $i = 0, 1, 2, \cdots, N_{iter}$, where $N_{iter}$ represents the number of FWI iterations, $\alpha_i > 0$ is the step length \cite{Nocedal_Wright_2006_Optimization_book}, and $h$ is the so-called descent direction (or search direction), which is linked to the gradient of the objective function, namely $\nabla_m \phi(m)$. In this work we employ the non-linear conjugate gradient method to minimize the objective function. In this approach, the search direction at the $i$-th iteration is given by \cite{Vigh2008}: 
\begin{equation}
        h(m_i) \!=\! 
  \begin{cases}
      \nabla_m \phi(m_0),& if \quad i = 0 \\
      \nabla_m \phi(m_i)\! +\!  \zeta (m_i) h(m_{i-1}),& \text{for} \,\, i = 1, 2, \cdots, N_{iter}
  \end{cases}
  \label{eq:searchdirection}
\end{equation}
with
\begin{equation}
    \zeta (m_i) = \frac{\nabla_m \phi(m_i)\big(\nabla_m \phi(m_i) - \nabla_m \phi(m_{i-1})\big)}{\nabla_m \phi(m_{i-1})\nabla_m \phi(m_{i-1})}.
\end{equation}

The gradient of the objective function \eqref{eq:FWI_lp_norm} with respect to the model parameters is given by the following expression:
\begin{equation}
    \nabla_m \phi_{L^p}(m) = \sum_{s,r} \int_{0}^{T} \Gamma_{s,r}^\dagger \frac{\partial \psi_s(m,t)}{\partial m}g_p \Big(\Delta d_{s,r}(m,t) \Big) dt,
    \label{eq:gradient_lp_norm}
\end{equation}
with
\begin{equation}
    g_p \Big(\Delta d_{s,r}(m,t) \Big) = \big|\Delta d_{s,r}(m,t)\big|^{p-1} \big[\text{sgn}\big(\Delta d_{s,r}(m,t)\big)\big]^p,
\end{equation}
where $\text{sgn}$ represents the sign function, $\Delta d_{s,r}(m,t) = \Gamma_{s,r} \psi_s(m,t) - d_{s,r}(t)$ is the residual data, $\frac{\partial \psi_s(m,t)}{\partial m}$ is the Fréchet derivative, and the superscript $\dagger$ refers to the transpose operator. Because calculating Fréchet derivatives is computationally expensive, the gradient of the objective function may be efficiently calculated using the adjoint-state method \cite{Plessix_GJI_AdjointMethodReview_2006}. In this approach the gradient of the objective function becomes:
\begin{equation}
    \nabla_m \phi_{L^p}(m) = -\sum_{s} \int_{0}^{T} q_s(T-t) \, \frac{\partial^2 \psi_s(t)}{\partial t^2} dt,
    \label{eq:gradient_lp_norm_adjoint}
\end{equation}
where $q_s$ is the backpropagated wavefield, the solution of the adjoint-state
equation:
\begin{equation}
\nabla^2 q_s(t) 
- m \frac{\partial^2 q_s (t)}{\partial  t^2 }   
= -\sum_r \Gamma_{s,r}^T g_p \Big(\Delta d_{s,r}(m,t) \Big),
\label{eq:adjoint_wave_equation_time}
\end{equation}
in which the spatial coordinate $\textbf{x}$ is implicit in the latter relations. 

The gradient of the objective function is computed efficiently by cross-correlating the forward wavefield $\psi_s$ with the backpropagated wavefield (or adjoint wavefield) $q_s$  in reverse time. In addition, we notice that the same FWI algorithm can be employed to compute the gradient of the $L^p$-based objective function since the only difference is the right-side term of Eq. \eqref{eq:adjoint_wave_equation_time}, namely the adjoint source. In the case $p=1$ the adjoint source is the sign function of the residual data \cite{Tarantola_1987_inverProblem_book,Crase_et_al_l1norm_Geophysics_1990}, and $p=2$ is the residual data ($\Gamma_{s,r} \psi_s - d_{s,r}$) \cite{Lailly_1983}.

\section{Brazilian pre-salt initial FWI setup} \label{sec:BrazilianpresaltinitialFWIsetup}

In this section we present the Brazilian pre-salt OBN dataset and the key features of the acoustic FWI workflow employed in this work. In this context, we introduce the details of our proposed FWI workflow to deal with refracted waves, which consists of an automatic data selection approach, data filtering and resampling, source estimation from the field data, source-receiver reciprocity, optimization of objective functions and gradient preconditioning.

\subsection{Circular shot OBN data} 

\begin{figure}[!htb]
\centering
\includegraphics[width=.4\columnwidth]{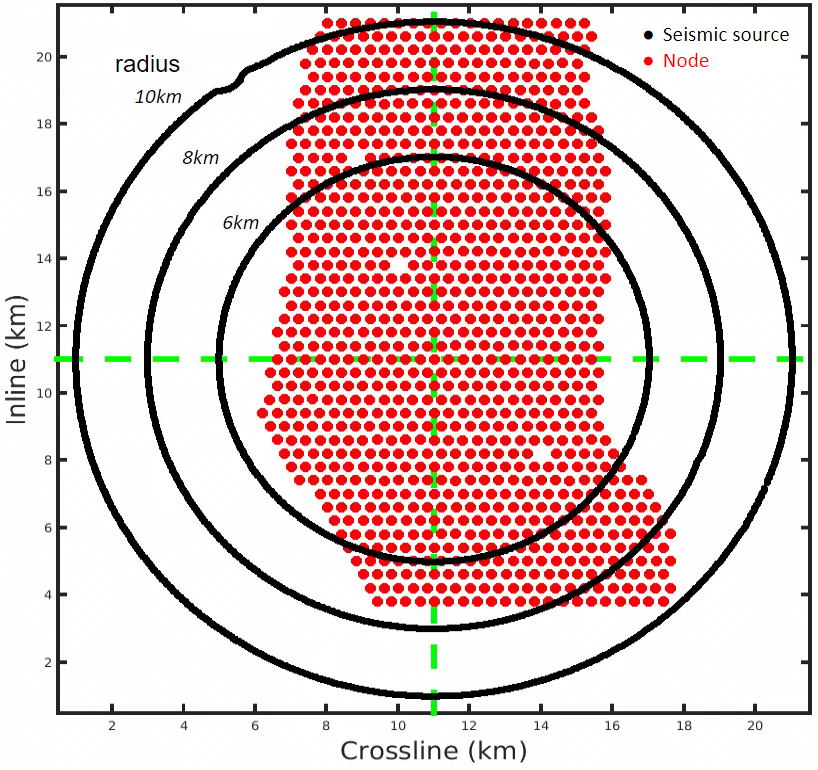}
\caption{Schematic map of the circular shot OBN survey, which comprises three concentric shooting circles (black curves) with radii of $6$, $8$ and $10km$, a total of $3006$ seismic sources, spaced every $50m$, and $1029$ four-component OBN (red dots) regularly spaced in a line interval of $400m$ ($200m$ staggered) over $44$ receiver lines. The dashed green lines indicate the central inline and central crossline. }
\label{fig:full_circular_shot_OBN_acquisition}
\end{figure}

The circular shot OBN survey was designed to maximize illumination of ultra-deep Brazilian pre-salt reservoirs by recording refracted waves at ultra-long offset and full azimuth \cite{Lopez_et_al_2020_SEG_OBNcircular}. This is an experimental dataset that consists of three concentric shooting circles with radii of $6$, $8$ and $10km$ with a regular shot spacing of $50m$, totaling $3006$ seismic sources. $1029$ 4C-OBN are distributed on the ocean floor ($2km$ depth average), regularly spaced in a line interval of $400m$ ($200m$ staggered) over $44$ receiver lines \cite{duarte_et_al_2023_Geophysics}. We depict a base map of the circular shot OBN geometry in Fig. \ref{fig:full_circular_shot_OBN_acquisition}, in which the black curves and the red points are the sources and nodes distribution. The three shot circles encompass $751$ ($6km$ circle), $1003$ ($8km$ circle) and $1252$ ($10km$ circle) seismic sources, respectively. 

\begin{figure*}[!htb]
     \centering
     \begin{subfigure}[b]{0.6\textwidth}
     \caption{}
     \label{fig:seismogram_original_a}
     \includegraphics[width=\textwidth]{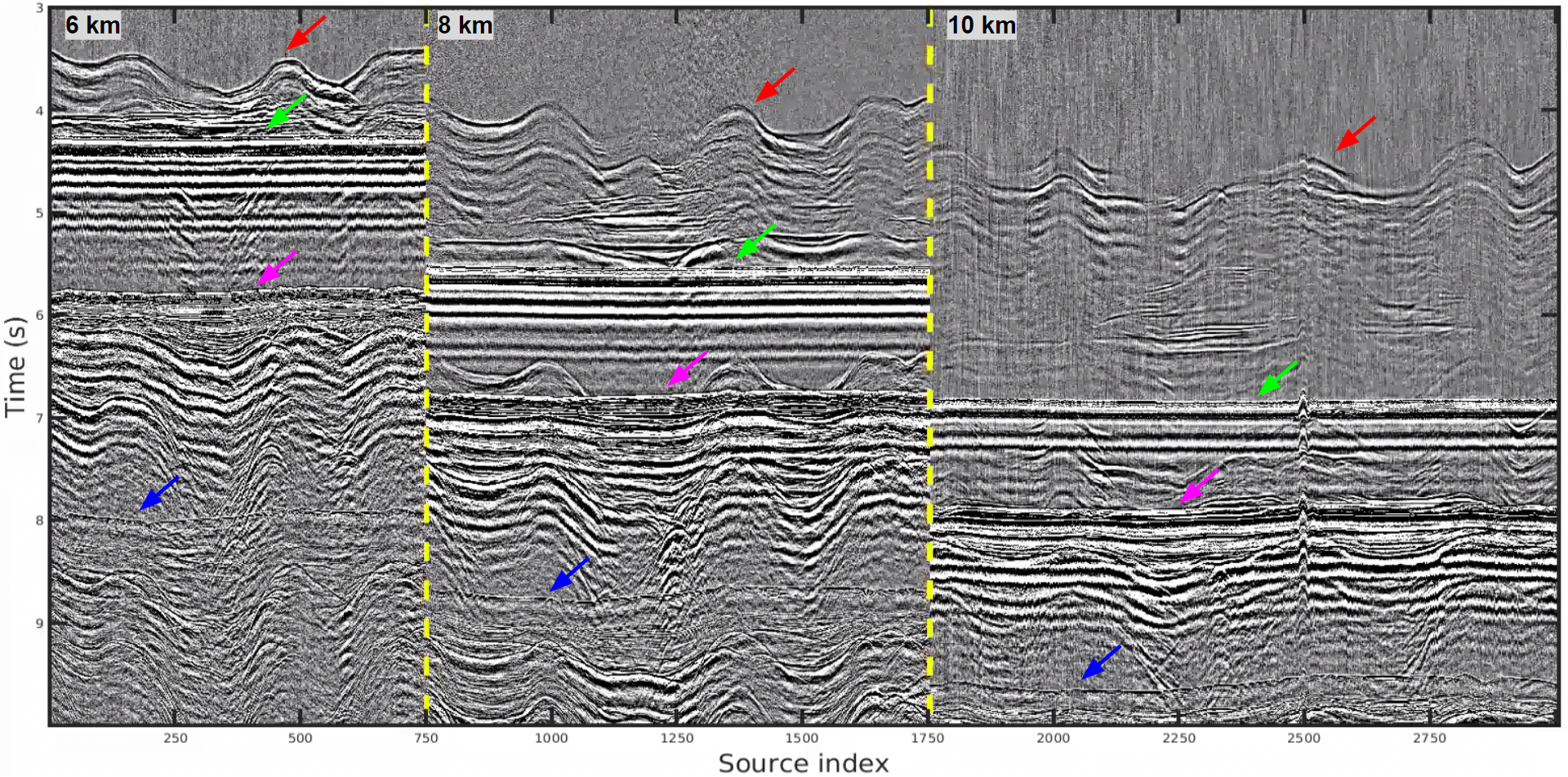}
     \end{subfigure}
       \\
     \begin{subfigure}[b]{0.6\textwidth}
     \label{fig:seismogram_original_b}
     \caption{}
     \includegraphics[width=\textwidth]{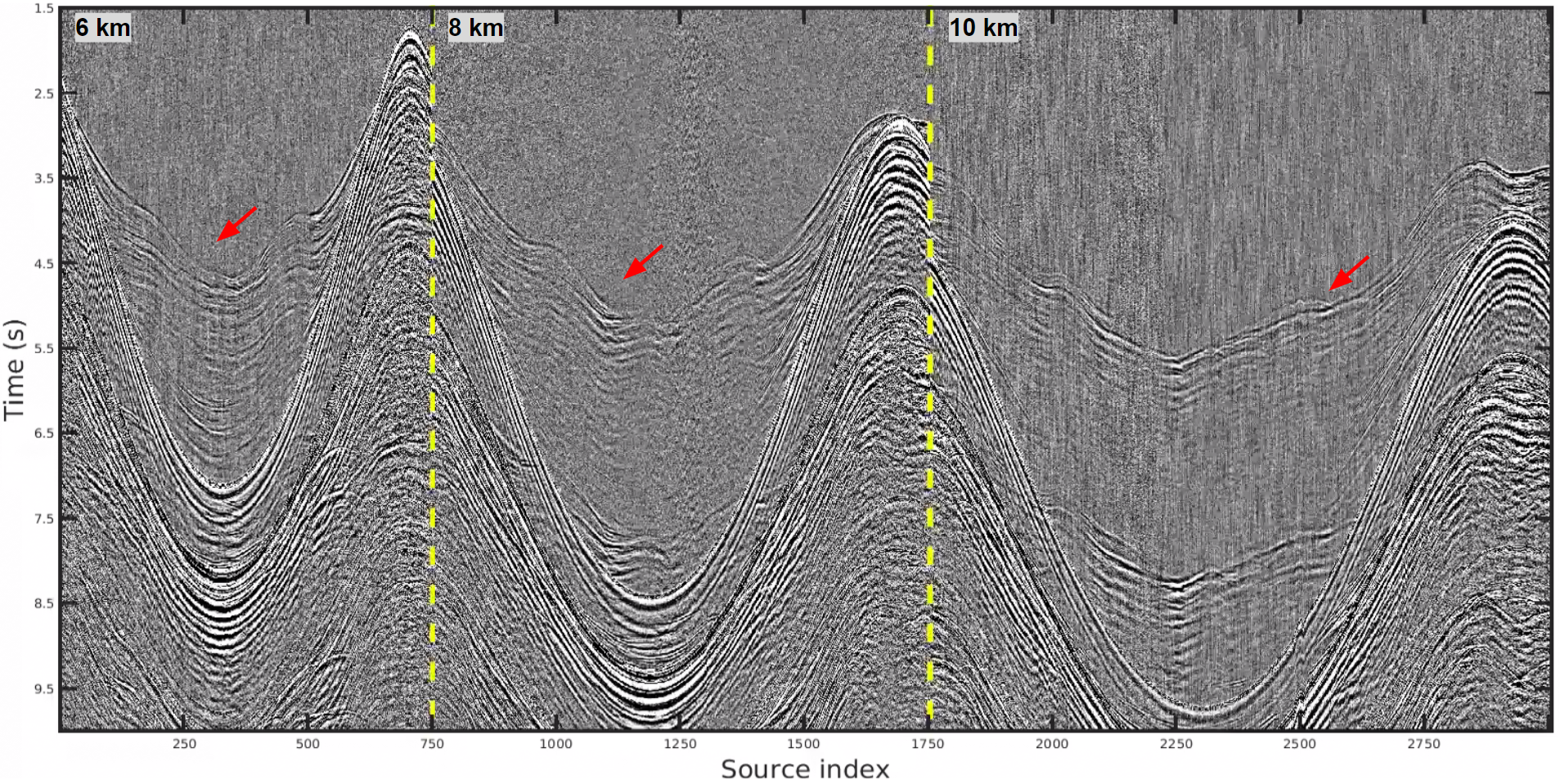}
     \end{subfigure}
     \caption{Common-receiver gathers, from the hydrophone component, for (a) the node at the center of the shot concentric circles (central node) and (b) the rightmost node of the central inline. In these gathers, the horizontal axis represents the traces from each source point along the corresponding circle. Since source points are equally spaced, the larger the radius, the more source points in the corresponding gather. The vertical dashed yellow lines indicate the limit between shot circles. The green, magenta, and blue arrows indicate the direct wave and its free air-water surface-related multiples, while the red arrows refer to the primary refractions.}
    \label{fig:seismogram_original}
\end{figure*}

In this work we consider only the hydrophone components. Figure \ref{fig:seismogram_original} shows two non-processed receiver gathers, in which panel (a) refers to the node at the center of the shot concentric circles (from now on, central node), while panel (b) refers to the rightmost node of the central inline. The receiver gathers consist of seismic traces grouped by shot circle and azimuth, which allows the visualization of receiver ''sub-gathers" linked to each shot circle (see yellow dashed lines in Fig. \ref{fig:seismogram_original}). The acquisition time is $10s$ with a $2ms$ sampling rate.

Analyzing the receiver-gathers in Fig. \ref{fig:seismogram_original}, we observe some interesting patterns. In the case of the central node (Fig. \ref{fig:seismogram_original}(a)), it is remarkable the intense flat events (same traveltime at same shot circle) associated with the direct wave in the water layer ($4.2s$, $5.5s$ and $6.8s$ for the circles with a radius of $6km$, $8km$ and $10km$, respectively). The free air-water surface-related multiples of the direct wave are also observed at times $5.8s$, $6.7s$ and $7.8s$ for the first multiple and $7.9s$, $8.7s$ and $9.5s$, for the following multiple, for the three circles, in which it is possible to note the slightly wavy pattern in contrast to the utterly flat as in the case of direct waves. Such an effect is due to the variation of the ocean bottom geometry by azimuth. As analyzed by Ref. \cite{Costa_et_al_2020_SEG_wavepaths}, the first and following immediate arrivals (red arrows in Fig. \ref{fig:seismogram_original}) refer to refractions from the top and base of salt and the pre-salt reservoirs. The red arrows in Fig. \ref{fig:seismogram_original}(b) show the registered refractions, demonstrating the importance of long offsets and full azimuth to record refracted waves. In this figure, we notice that the refracted waves are quickly identified, which reinforces the potential of this acquisition geometry to provide refracted waves that penetrate ultra-deep regions, favoring a robust pre-salt velocity reconstruction \cite{daSilva_et_al_SEG_2021_qFWI_BrazilianPreSalt}. Circular shot OBN acquisition clearly provides high-quality refracted waves recording from long offsets. Such waves travel through the deep interior of the subsurface, illuminating the Brazilian pre-salt reservoirs.

\subsection{Initial model}

\begin{figure}[!htb]
    \centering
    \includegraphics[width=.5\columnwidth]{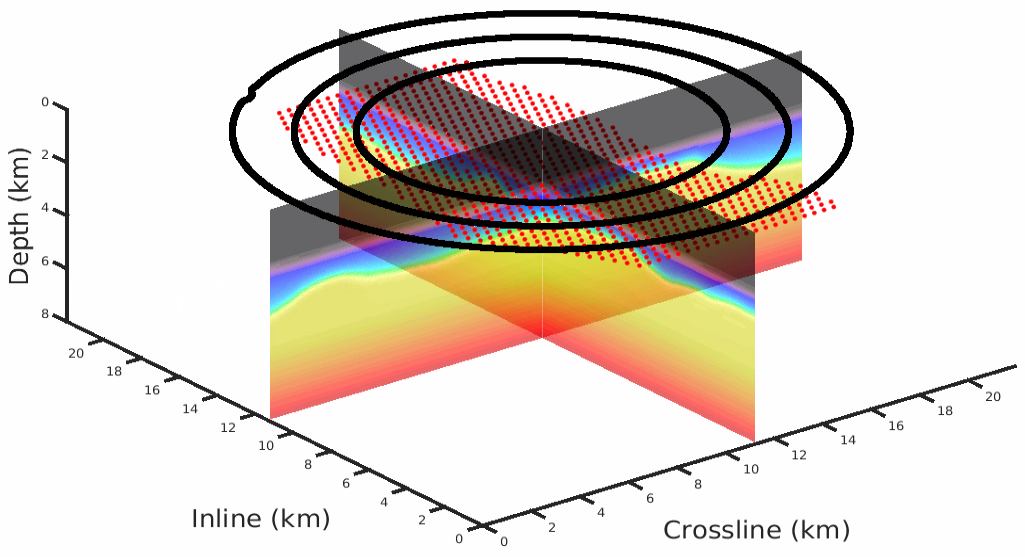}
    \caption{3D perspective of the initial FWI model and the circular shot OBN acquisition geometry. The studied area consists of a parallelepiped region of $22km \times 22km \times 8km$. Notice that the pre-salt region (depth $\approx 6$--$8km$) is a 1D velocity model that linearly increases at depth, discarding previous information from the pre-salt reservoirs.}
\label{fig:initial_model_acquistion_3Dfigure}
\end{figure}

In this work we construct the initial model $m_0$ by smoothing the model representing the Brazilian pre-salt presented in Refs. \cite{daSilva_et_al_SEG_2021_qFWI_BrazilianPreSalt,daSilva_et_al_2022_EAGE_KleinGordon} but without the pre-salt region. In this regard we substitute the pre-salt area (depth $>6km$) with a 1D model, which increases with depth according to a linear function. The idea is to discard any prior information from the pre-salt region to assess the effectiveness of the refraction FWI using the circular shot OBN geometry. We use a Gaussian filter with a standard deviation of $240m$ for the smoothing model. Figure \ref{fig:initial_model_acquistion_3Dfigure} shows the initial model and the circular shot OBN acquisition, in which the studied area consists of a parallelepiped region of $22km \times 22km \times 8km$.

\subsection{Data preparation and automatic selection \label{sec:Data_preparation_automatic_selection}}

Since refracted waves are essential for recovering the long-wavelength structure of the subsurface model \cite{Kazei_et_al_2013_FWI}, we perform a data selection before the FWI process to capture such seismic phases. In particular, we execute an automatic data selection using the Eikonal equation for scalar waves, given by \cite{Noble_et_al_2014_GJI_Eikonal}:
\begin{equation}
\nabla \tau \cdot \nabla \tau = \frac{1}{v_p^2(\textbf{x})},
\label{eq:eikonal}
\end{equation}
where $\tau$ is the traveltime of the first arrival. This equation gives information about the kinematic features of wave propagation (wavefront).  We use the initial velocity model depicted in Fig. \ref{fig:initial_model_acquistion_3Dfigure} and Eq. \eqref{eq:eikonal} to determine a time window for each seismic trace.  We consider the original and mirrored node positions to obtain upgoing and downgoing (free-surface-related multiples) first arrival waves used for constructing a time window that confines refracted waves in seismograms. Traveltimes associated with the upgoing first arrivals are used as the lower bound of the time window, while traveltimes related to the downgoing first arrivals are employed as the upper bound. It is worth noting that the first arrivals are not always refractions; in most cases, they are, as analyzed in Ref. \cite{Costa_et_al_2020_SEG_wavepaths}.

\begin{figure}[!htb]
     \centering
     \begin{subfigure}[b]{0.481\textwidth}
     \caption{}
     \includegraphics[width=\columnwidth]{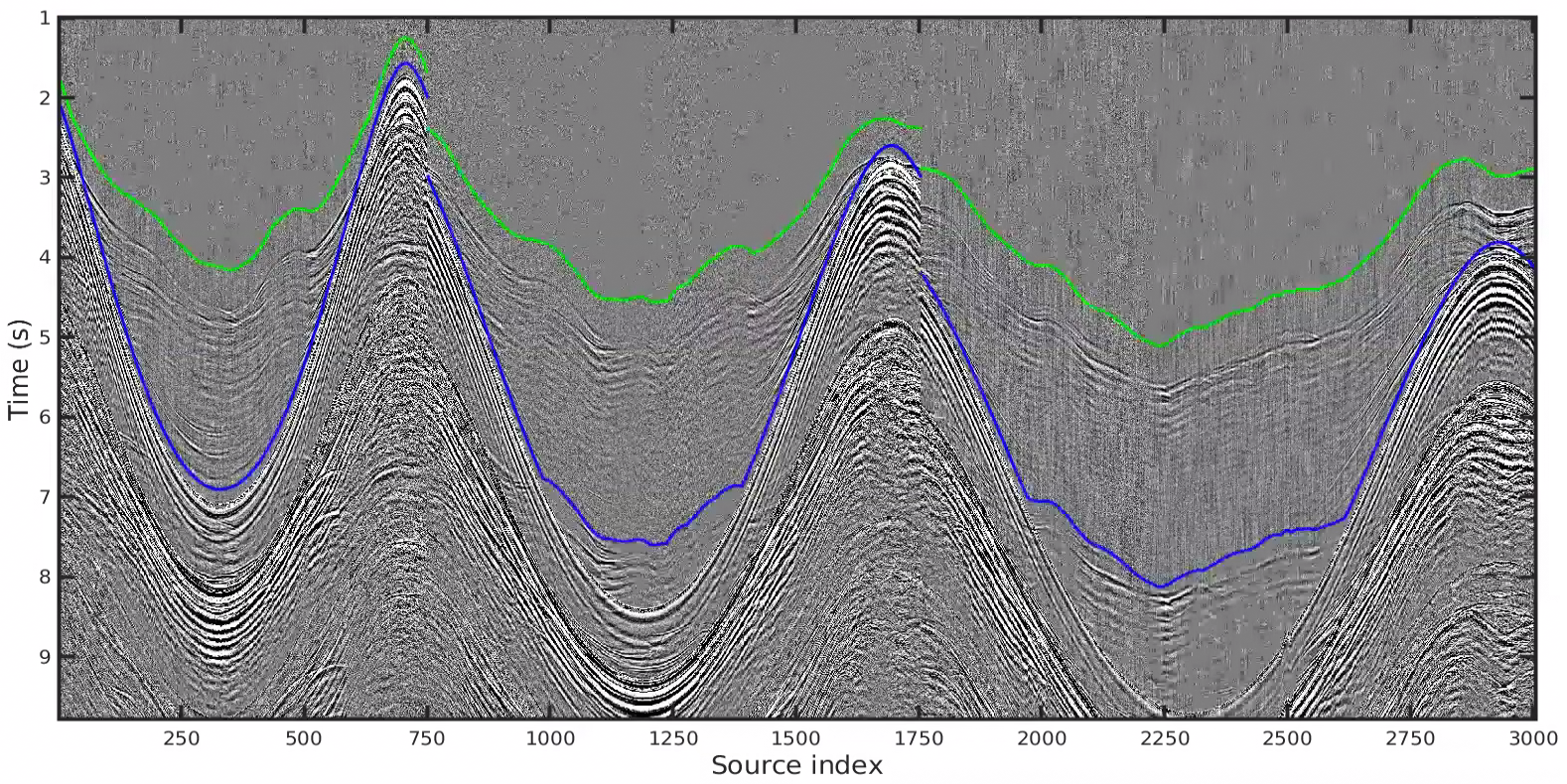}
     \label{fig:seismogram_original_muted}
     \end{subfigure}
       \\
    \begin{subfigure}[b]{0.481\textwidth}
     \caption{}
     \includegraphics[width=\columnwidth]{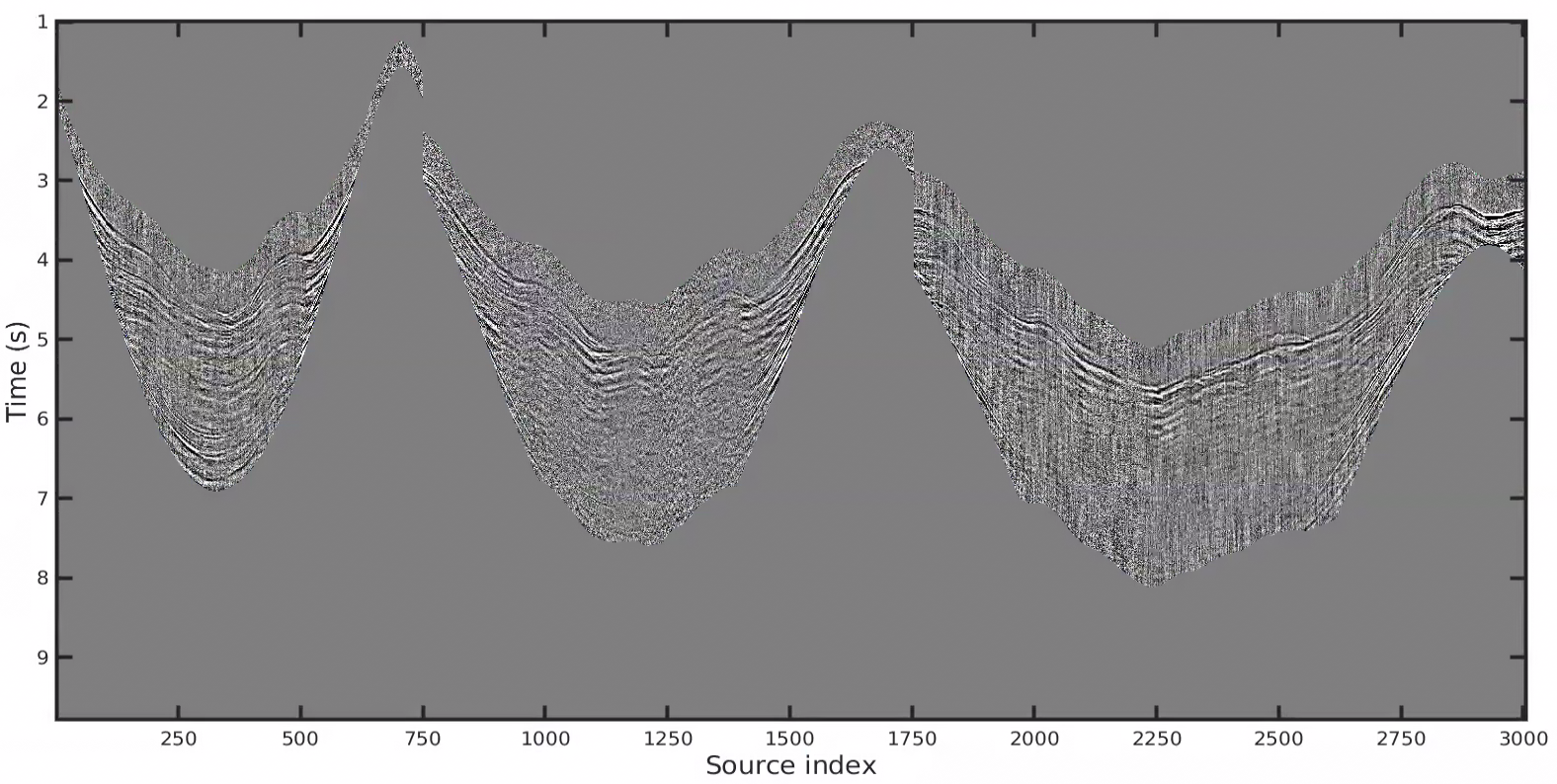}
     \label{fig:seismogram_original_mutedb}
     \end{subfigure}
       \\
       \begin{subfigure}[b]{0.481\textwidth}
     \caption{}
     \includegraphics[width=\columnwidth]{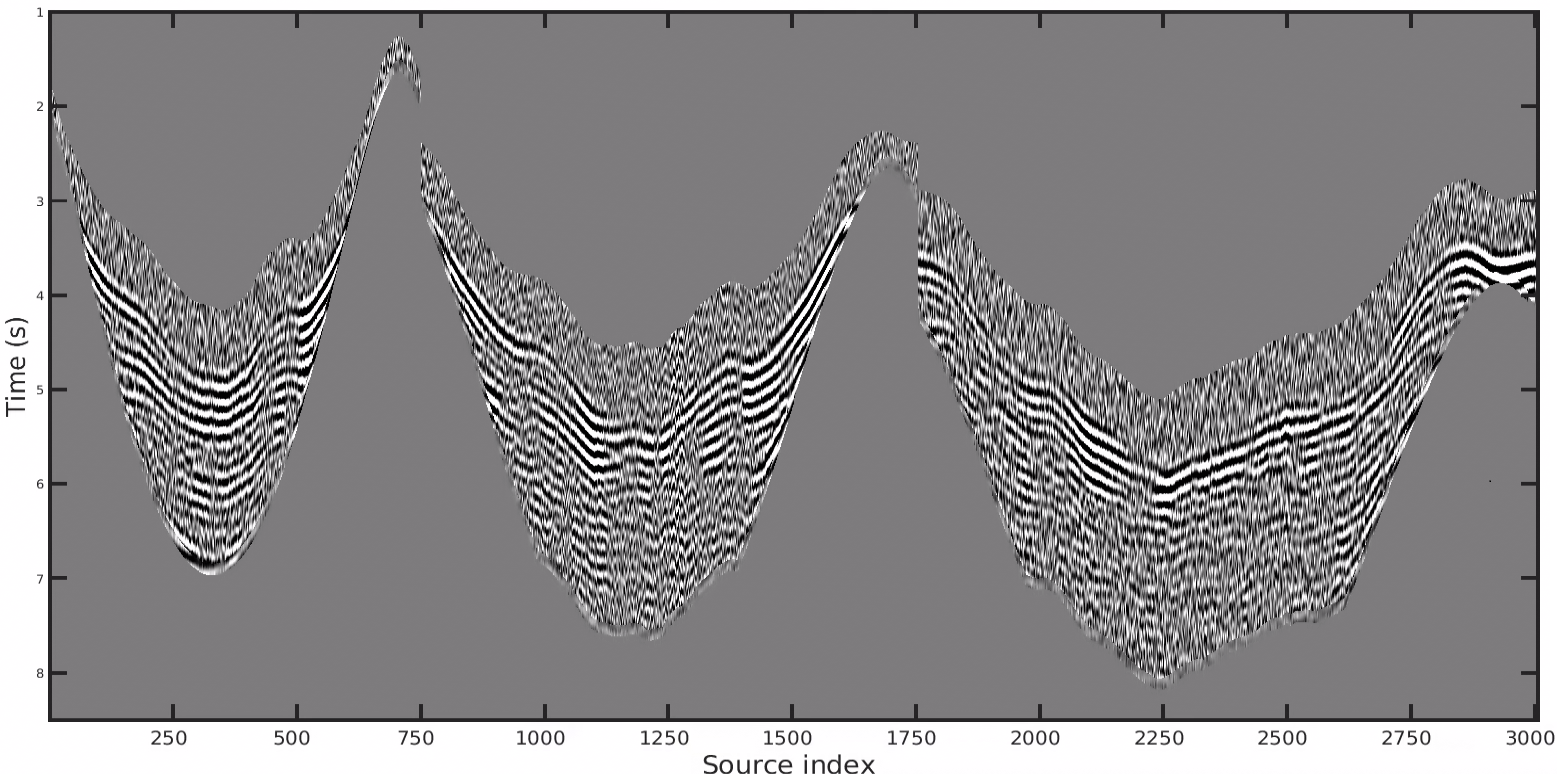}
     \label{fig:seismogram_filtered_muted}
     \end{subfigure}
       
     \caption{(a) Time window computed by the Eikonal equation \eqref{eq:eikonal} for the common receiver-gather from the hydrophone component of the rightmost node of the central inline. Green and blue lines represent the lower and the upper time limits. Waveforms recorded before the green line or after the blue line are muted, as depicted in panel (b). Panel (c) shows a common receiver-gather from the hydrophone component for the rightmost node of the central inline after filtering and muting. The times used for muting the receiver gather are the same as shown in Panel (a), while the filter consists of an Ormsby bandpass with corner frequencies of $1$–$2$–$4$–$5$Hz.}.
    \label{fig:seismogram_original_muting}
\end{figure}

\begin{figure}[!htb]
     \centering
     \begin{subfigure}[b]{0.31\textwidth}
     \caption{}
     \includegraphics[width=\columnwidth]{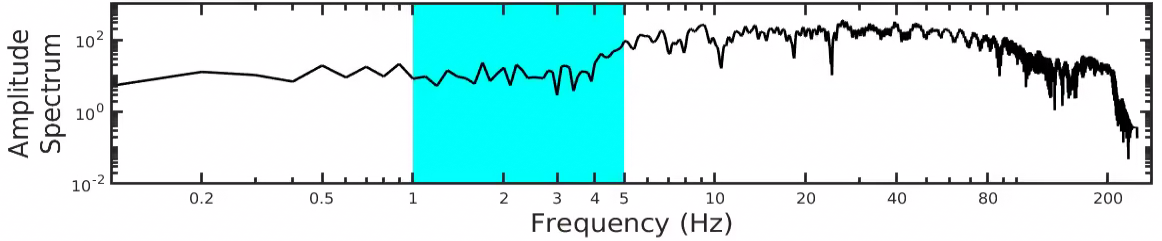}
     \label{fig:seismogram_spectruma}
     \end{subfigure}
       \hfill
       \begin{subfigure}[b]{0.31\textwidth}
     \caption{}
     \includegraphics[width=\columnwidth]{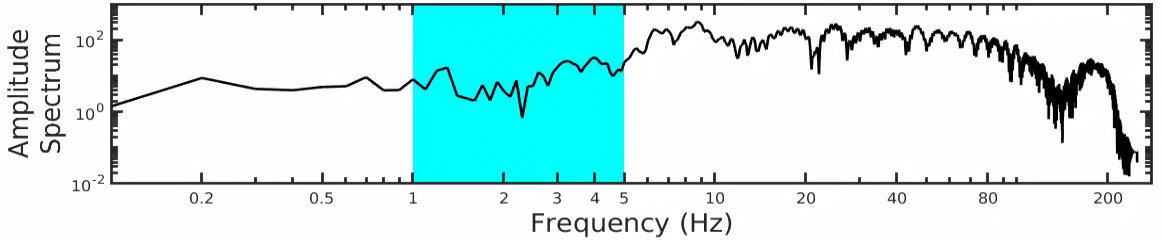}
     \label{fig:seismogram_spectrumb}
     \end{subfigure}
       \hfill
     \begin{subfigure}[b]{0.31\textwidth}
     \caption{}
     \includegraphics[width=\columnwidth]{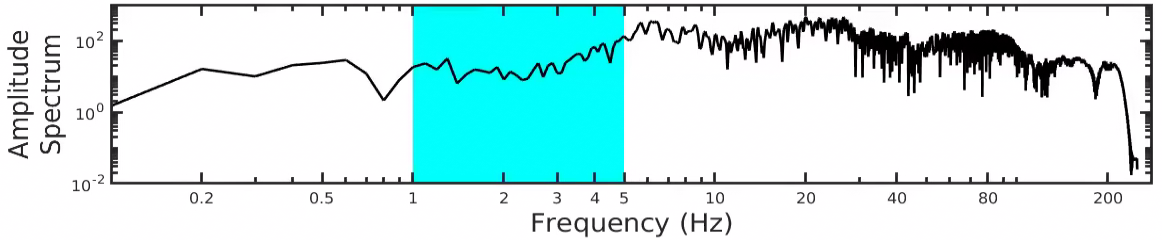}
     \label{fig:seismogram_spectrumc}
     \end{subfigure}
       \hfill
     \caption{Amplitude spectra of seismic data from traces with source indexes (a) 250, (b) 1000, and (c) 2750, in Figs. \ref{fig:seismogram_original_muted} and \ref{fig:seismogram_original_mutedb}. The cyan rectangle denotes the considered spectrum region in this study. }
    \label{fig:seismogram_spectrum}
\end{figure}

Due to the inaccuracies of the initial model, the first arrivals calculated with the Eikonal equation naturally do not match the field data perfectly. In this way, we shift the calculated traveltimes by $-0.5s$ and $-0.2s$, respectively, for the original nodes and mirrored nodes cases. We apply these shifts to all receiver gathers to ensure the consistency of the entire dataset. Then, we mute all signal amplitudes with times less than the lower limit or greater than the upper limit. We apply a Gaussian smoothing filter to remove possible high contrasts at the time limits. Figure \ref{fig:seismogram_original_muted} shows the common receiver-gather from the hydrophone component of the rightmost node of the central inline, in which the green and blue lines represent the lower and the upper time limit computed by the Eikonal equation \eqref{eq:eikonal}. Figure \ref{fig:seismogram_original_mutedb} displays the identical receiver gather after the muting process. Figure \ref{fig:seismogram_spectrum} displays the seismic data's amplitude spectra, which correspond to the seismic traces of the receiver-gather depicted in Figs. \ref{fig:seismogram_original_muted} and \ref{fig:seismogram_original_mutedb}, with source indexes 250, 1000, and 2750, respectively. The cyan rectangle highlights the spectrum region being studied.

After data selection, we apply an Ormsby bandpass filter with corner frequencies $f_1$-$f_2$-$f_3$-$f_4Hz$ to perform the low-frequency FWI,  where $f_1$ and $f_4$ are low- and high-cut frequencies and $f_2$ and $f_3$ are low- and high-pass frequencies, respectively. Moreover, to speed up the forward modeling and the objective function gradient calculation, we perform a data resampling according to the Nyquist–Shannon sampling theorem. In this work we consider $f_1 = 1$, $f_2 = 2$, $f_3 = 4$ and $f_4 = 5Hz$. Figure \ref{fig:seismogram_filtered_muted} shows the common receiver-gather from the hydrophone component of the rightmost node of the central inline after the filter and mute application.

\subsection{Source estimation \label{sec:source_estimation}}

To obtain an accurate source signature estimation, we employ the methodology proposed in Ref. \cite{Pratt_1999_SourceEstimation}. In this approach, the source wavelet is estimated in the frequency domain, using all frequencies within the range of the spectral content of the filtered field data in the least-squares sense, as follows:
\begin{equation}
 S(\omega) = \frac{\sum_i G^\dagger_i(\omega)D^*_i(\omega)}{\sum_i G_i^\dagger(\omega)G_i^*(\omega)},
\label{eq:source_estimation}
\end{equation}
where $S$ is the frequency-domain source wavelet, $G$ and $D$ are, respectively, Green's function and field data in the frequency domain, $\omega$ denotes the angular frequency and the superscripts $\dagger$ and $*$ represent, respectively, the complex conjugate and the transposition operator.

We select $22$ seismic traces with the shortest source-receiver distance from the filtered observed data in this work. Then, we compute Green's function in the time domain by propagating a Delta Dirac function on the initial model using the acquisition geometry associated with the previously selected traces and the same forward modeling operator that will be used for the FWI procedure. Next, we calculate the Fourier transform of the selected data and the Green's function to estimate the source signature using Eq. \eqref{eq:source_estimation}. Finally, we compute the inverse Fourier transform of  $S(\omega)$ to obtain the source signature in the time domain. Since we only applied a band-pass filter on field data and estimated the source wavelet directly from them, we assumed that ghost and bubble effects will be simulated by our forward modeling algorithm. It is also worth highlighting that the Green's function possesses energy distributed uniformly across all frequencies due to its flat and unlimited frequency spectrum. In this way, we constrain the source estimation process using Eq. \eqref{eq:source_estimation} in the frequency range of the observed data set to comply with the Nyquist-Shannon theorem. Thus, in accordance with the Nyquist-Shannon theorem, we estimate the seismic source at frequencies lower than the Nyquist frequency by considering the highest frequency present in the filtered data.

\begin{figure}[!htb]
\centering
\flushleft{\hspace{.1cm} (a) \hspace{7.6cm} (b)}
\includegraphics[width=.95\columnwidth]{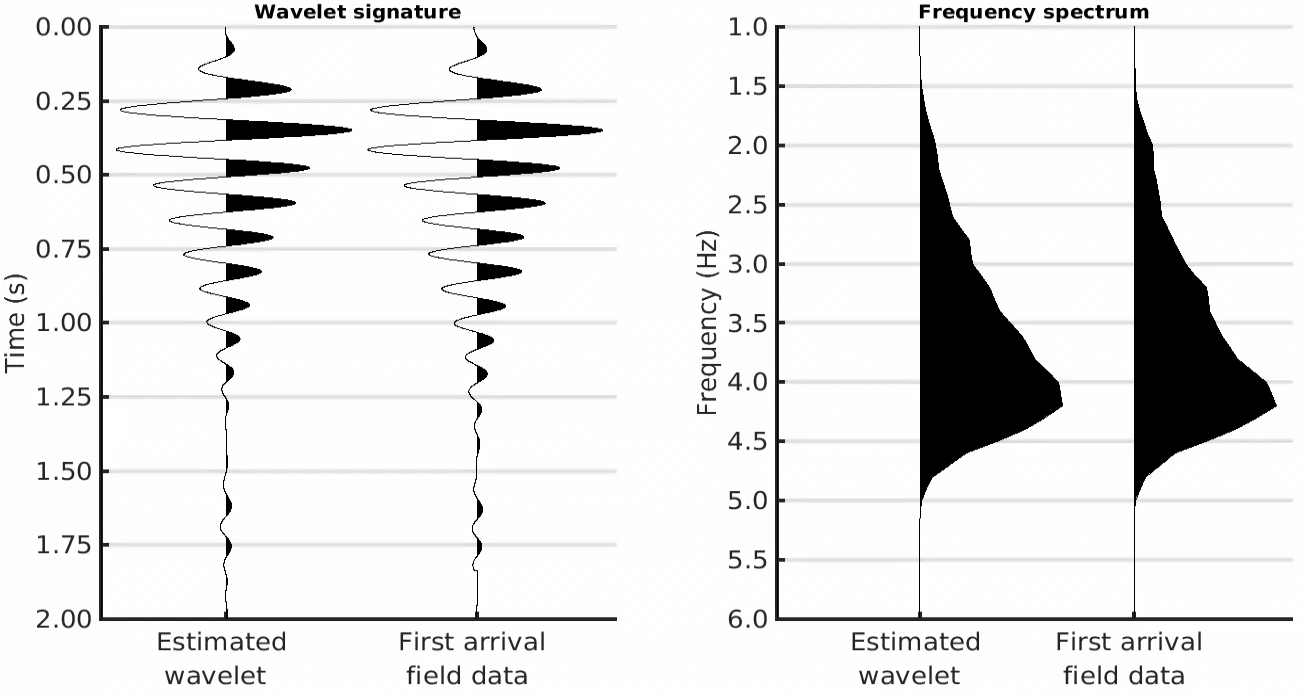}
\caption{Estimated source wavelet and first arrival extracted from the field dataset (corner frequencies $1$–$2$–$4$–$5$Hz) in (a) time domain, and (b) frequency domain. The similarity between them is remarkable.}
\label{fig:source_wavelet_estimated_pratt1999}
\end{figure}

Figure \ref{fig:source_wavelet_estimated_pratt1999} shows the estimated source wavelet from the filtered field data (corner frequencies $1$–$2$–$4$–$5$Hz) and the first arrival recorded in the field dataset. We notice that the waveforms are very similar, including the bubble effect around $\approx 1.50-1.75$s, as depicted in Fig. \ref{fig:source_wavelet_estimated_pratt1999}(a). Figure \ref{fig:source_wavelet_estimated_pratt1999}(b) shows the amplitude spectrum of the wavelets presented in Fig. \ref{fig:source_wavelet_estimated_pratt1999}(a), in which the similarity between them is remarkable.

\subsection{Forward modeling and gradient computation} 

Our modeling (and inversion) code, called \textit{SEISWAVE3D}, is implemented in the C++ Programming Language by employing the massive parallelism of GPUs using the OpenACC routine directive \cite{Farber_openACC_nook2016}. In this regard we generate the modeled waveforms by solving Eq. \eqref{eq:wave_equaqtion_time} employing a 3D finite-difference method ($2nd$ order in time and $8th$ order in space). We impose absorbing sponge layers around the model to simulate an infinite medium \cite{Cerjan_et_al_1985_Geophysics_Bordas}. Furthermore, since the true positions of sources and receivers (nodes) may not match the  finite-difference grid point, we accurately position sources and receivers using Kaiser windowed sinc functions \cite{Hicks_2002_Geophysics_interpolation}.

Let us remind the reader that the high computational requirements of FWI algorithms are primarily due to gradient computation \eqref{eq:gradient_lp_norm_adjoint}, which can be broken down into three steps: (i) the forward modeling for obtaining the modeled data \eqref{eq:wave_equaqtion_time}; (ii) the backpropagating modeling for calculating the adjoint wavefield \eqref{eq:adjoint_wave_equation_time}; and (iii) the application of an imaging condition by cross-correlating the forward wavefield with the adjoint wavefield. Such steps spend most of the FWI runtime. In addition, the gradient computation is memory-consuming and impractical depending on the size of the model \cite{Symes_2007_RTM_OptimalCheckPoint}. Thus, we compute the gradient by applying an approach that balances runtime and computational memory consumption proposed by Ref. \cite{Clapp_2009_SEG_RandonBoundaries}. In this framework, we compute the adjoint equation using a reverse time migration (RTM) algorithm that consists of backpropagating the adjoint source using the two last snapshots of the forward wavefield and random boundaries \cite{Clapp_2012_SEG_RandonBoundaries}. This approach significantly reduces memory and IO requirements compared to other techniques \cite{Salamanca_et_al_2018}, allowing large-scale 3D FWI to be performed on a single GPU. Furthermore, we resample the filtered data accordingly to the Nyquist-Shannon sampling theorem to speed up the forward modeling and objective function gradient calculation. In particular, we subsampled the original data from $2ms$ to $6ms$, decreasing the runtime of computing a single gradient and saving GPU memory, as will be discussed further. 

In this work we consider four different scenarios regarding the gradient preconditioning used in the FWI process. The first is a simple case in which the gradient is free from previous conditioning. We consider a smoothed gradient in the second one using an anisotropic nonstationary Bessel filter \cite{Trinh_et_al_2017_BesselFilter}. Third, we employ a preconditioned gradient using source-receiver illumination \cite{Kaelin_2007_EAGE}. In the last, we consider a combination of second and third scenarios. The idea is to check which scenario is most suitable to mitigate possible acquisition footprints in the refraction FWI results.

\subsection{FWI workflow}

In this section we summarize our FWI strategy and discuss some additional points. The pivotal elements of the FWI workflow to deal with refracted waves are related to the acquisition geometry, data selection and filtering, source wavelet estimation, and gradient computation, which may be summarized as follows:
\begin{description}
    \item[(i)] \textbf{Automatic data selection} using the Eikonal equation for extracting the waveforms from the dataset by selecting diving waves recorded before the direct wave or first arrivals free-surface-related multiples from the surface/air interface;
    \item[(ii)] \textbf{Data filtering} using Ormsby bandpass filter with corner frequencies $f_1$-$f_2$-$f_3$-$f_4$ Hz;
    \item[(iii)] \textbf{Source estimation} from the filtered data recorded at the smallest source-receiver distance, using the Green's function computed with our forward modeling operator;
    \item[(iv)] \textbf{Data resampling} accordingly to the Nyquist–Shannon sampling theorem to speed up forward modeling and objective function gradient calculation. This strategy allows decreasing the number of time steps of the numerical solution of the wave equation without loss of information, as discussed, for instance, by Ref. \cite{Yang_et_al_2016_Geophysics_81_resamplingGradient};
    \item[(v)] \textbf{Random shot sampling } by randomly choosing seismic sources to save computational efforts and to avoid regular interference patterns on the gradient \cite{Diaz_Guitton_2011_SEG_RandomShotDecimation,Warner_et_al_2013_Geophysics_AnisotropicFWI,Kamath_et_al_2021_Geophysics_FWI_Multiparameter};    
    \item[(vi)] \textbf{Reciprocity principle}: Interchanging source and receiver locations in the FWI to reduce computational efforts. This step is only beneficial to be applied when the number of receivers is smaller than the number of seismic sources, which is our case;
    \item[(vii)] \textbf{Nonlinear optimization algorithm} for minimizing an objective function. In our case, we choose a nonlinear conjugate gradient method;
    \item[(viii)] \textbf{Gradient preconditioning} using source-receiver illumination \cite{Kaelin_2007_EAGE} and the anisotropic nonstationary Bessel filter \cite{Trinh_et_al_2017_BesselFilter}.
\end{description}

The previously presented workflow is efficient in both computational and operational terms. For instance, points (i)-(vi) are fundamental for saving work time in the seismic industry, especially the first one. Point (vii) is a topical scientific research field with numerous formulations, each with its advantages and disadvantages.

\section{Brazilian pre-salt case study} \label{sec:results}

In this section we demonstrate the application of our proposed FWI workflow to the Brazilian pre-salt circular shot OBN dataset to estimate P-wave velocities. In particular, we present two applications: (i) in the first, called "narrow test", we focus our study around the central crossline by using a subset of the circular shot OBN survey distributed in a region of $22km \times 5km \times 8km$. (ii) In the second, called "full azimuth test", we take into account the full azimuthal coverage offered by the circular shot OBN dataset. The narrow test allows us to perform inversion tests more quickly, allowing the validation of our proposed workflow for employing in the full azimuth test.

In our analyses we consider only the hydrophone components in the $2$--$5Hz$ frequency band to be consistent with the acoustic FWI modeling engine. This low-frequency analysis allows us to use a relatively large grid spacing, reducing the computational cost \cite{Bunks_et_al_Multiscale_Geophysics_1995}. We discretize the initial model (Fig. \ref{fig:initial_model_acquistion_3Dfigure}) on a regular grid with a spacing of $80m$. To reduce the computational efforts, we applied the reciprocity principle by interchanging source and receiver locations in the FWI process, labeling receivers as sources and sources as receivers. In addition, we resample the waveforms from $2ms$ to $6ms$ to speed up the forward modeling and the objective function gradient calculation. To compute the gradient, we consider subsets of nodes with good seismic coverage in the receiver domain instead of using all receiver gather, which is an efficient strategy, as discussed in several works  \cite{Diaz_Guitton_2011_SEG_RandomShotDecimation,Warner_et_al_2013_Geophysics_AnisotropicFWI,vanLeeuwen_Herrmann_GeophysProspecting_2013,Kamath_et_al_2021_Geophysics_FWI_Multiparameter,Cobo_et_al_segam2021_DecimationSourcesOBN_SantosBasin}. We also include a simple mask in the gradient (from $0$ to $2km$ depth) for muting the water layer during the inversion process.

\subsection{FWI around central crossline (narrow test)}

\begin{figure*}[!htb]
\flushleft{\hspace{2.2cm}(a) \hspace{4.5cm} (b) \\}
\centering
\includegraphics[width=.6\textwidth]{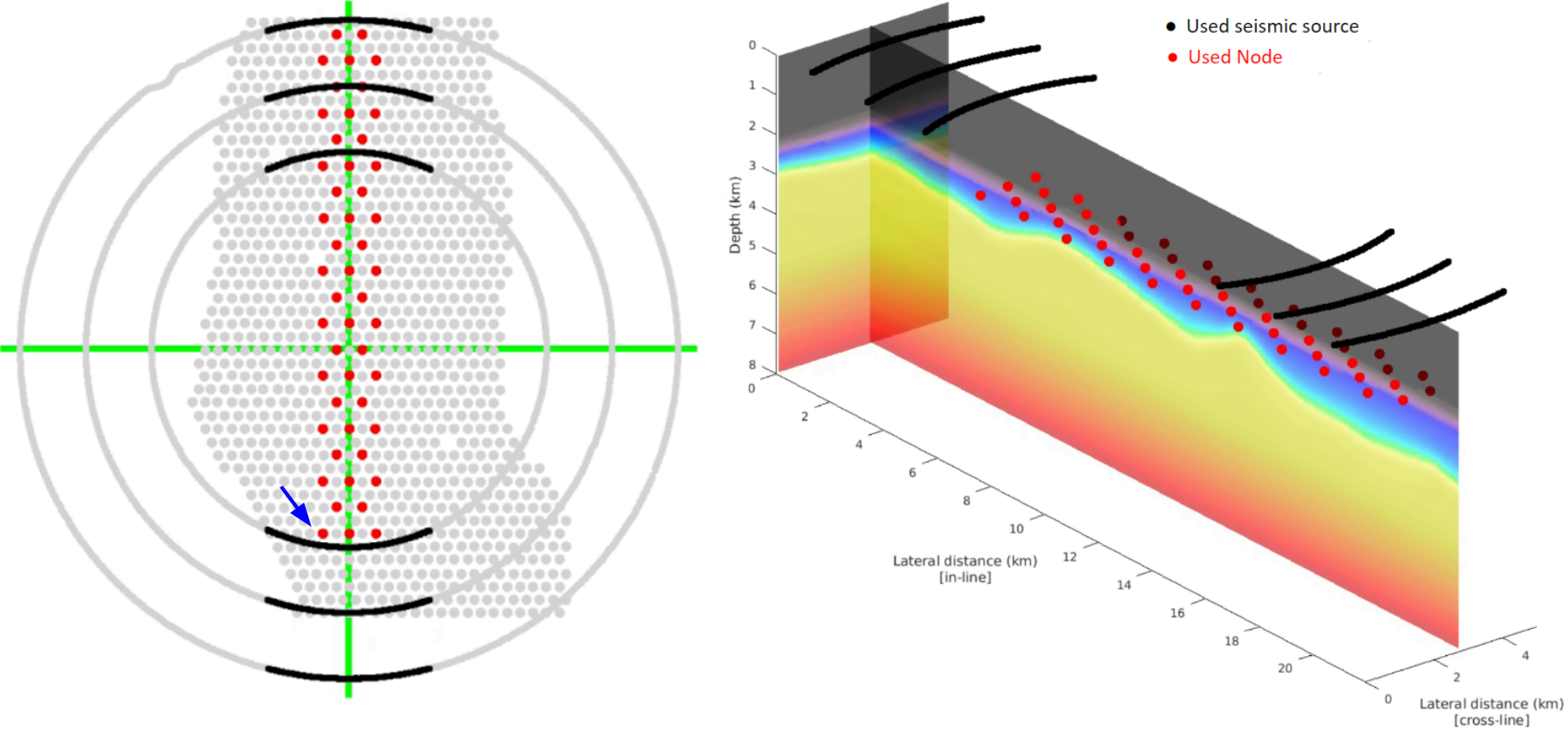}
\flushleft{\vspace{-.8cm}\hspace{2.2cm}(c) \\}
\centering
\includegraphics[width=.65\textwidth]{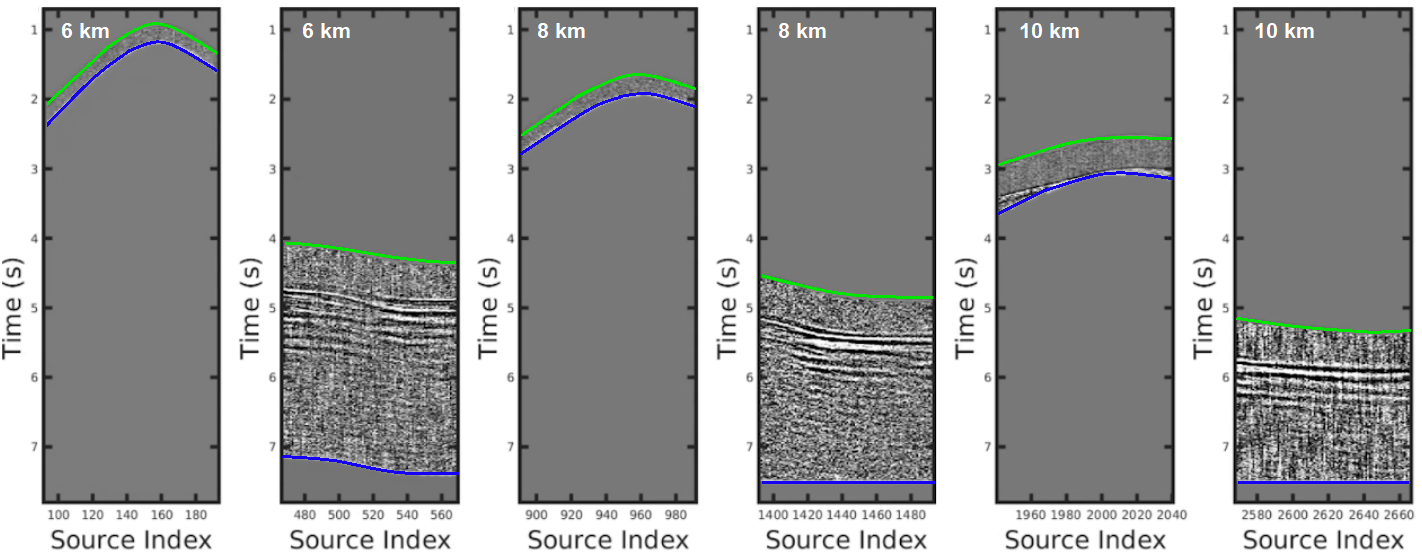}
\caption{(a) Base map of the circular shot OBN survey around the central crossline used in the "narrow test", which comprises $607$ seismic sources (black curve), and $50$ nodes (red dots). The green lines indicate the central inline and central crossline. (b) 3D perspective of the initial FWI model and the circular shot OBN acquisition geometry depicted in Panel (a), comprising a parallelepiped region of $22km \times 5km \times 8km$. (c) Example of selected receiver gather used in the "narrow test”, referring to the node indicated by the blue arrow in Panel (a). Green and blue lines represent the lower and the upper time limit computed by Eikonal solution. Waveforms recorded before the green line and after the blue line are muted. 
}
\label{fig:geometry_initial_model_CrossLine_NarrowTest}
\end{figure*}

To assess the effectiveness of our proposed FWI workflow for dealing with refracted waves, we first focused our study around the central crossline ("narrow test"). In this regard we consider a subset of the circular shot OBN survey (Fig. \ref{fig:full_circular_shot_OBN_acquisition}), comprising  $607$ seismic sources and $50$ nodes distributed in a parallelepiped region of $22km \times 5km \times 8km$ (see Figs. \ref{fig:geometry_initial_model_CrossLine_NarrowTest}(a) and \ref{fig:geometry_initial_model_CrossLine_NarrowTest}(b)). Figure \ref{fig:geometry_initial_model_CrossLine_NarrowTest}(c) shows an example of receiver gather used in this test, which is muted according to the Eikonal equation solution as discussed in the previous section. We discretize the initial model (Fig. \ref{fig:geometry_initial_model_CrossLine_NarrowTest}(b)) on a regular grid with a spacing of $80m$, with a total of $7,693,776$ grid points ($276 \times  276 \times  101$ cells).
 
We perform 3D time-domain FWI by employing the objective functions based on $L^1$ and $L^2$ criteria, in which we set up $10$ iterations for all data inversions. We also consider the four gradient preconditioning scenarios: (i) gradient without preconditioning; (ii) preconditioned gradient using source-receiver illumination; (iii) smoothed gradient using an anisotropic nonstationary Bessel filter; and (iv) a combination of scenarios (ii) and (iii). For all inversion processes, we use a computer with a Dell Precision 3930 Rack, 6 cores, 128GB ram, NVIDIA Quadro A2000 12GB. Regardless of the gradient preconditioning scenario, each FWI simulation takes about 12 hours. 
 
\begin{figure}[!htb]
\flushleft{\hspace{4cm}(a) \\}
\centering
\includegraphics[width=.5\textwidth]{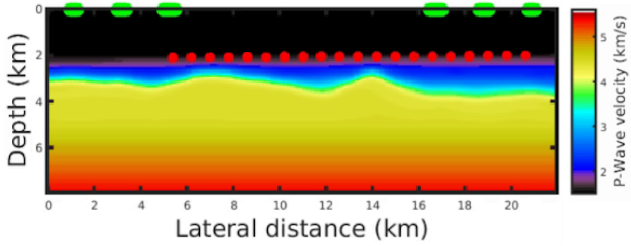}
\flushleft{\vspace{-.25cm}(b) \hspace{8cm} (c)}
\includegraphics[width=\textwidth]{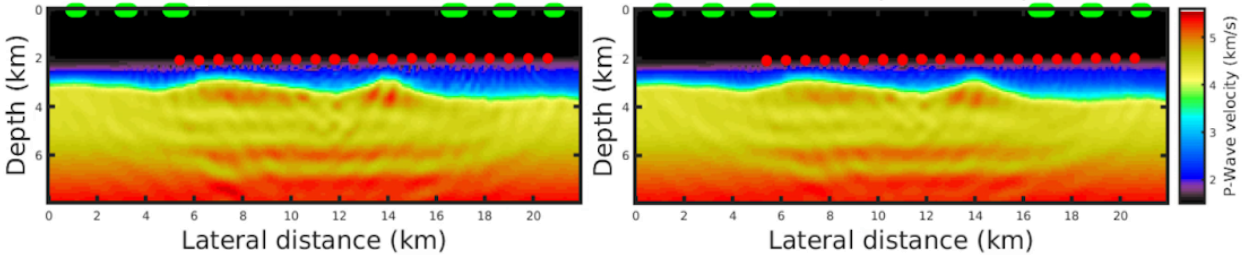}
\flushleft{\vspace{-.3cm}(d) \hspace{8cm} (e)}
\includegraphics[width=\textwidth]{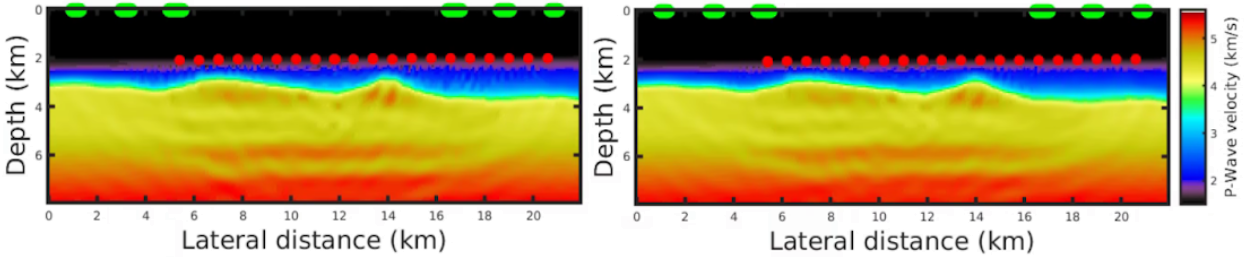}
\flushleft{\vspace{-.3cm}(f) \hspace{8cm} (g)}
\includegraphics[width=\textwidth]{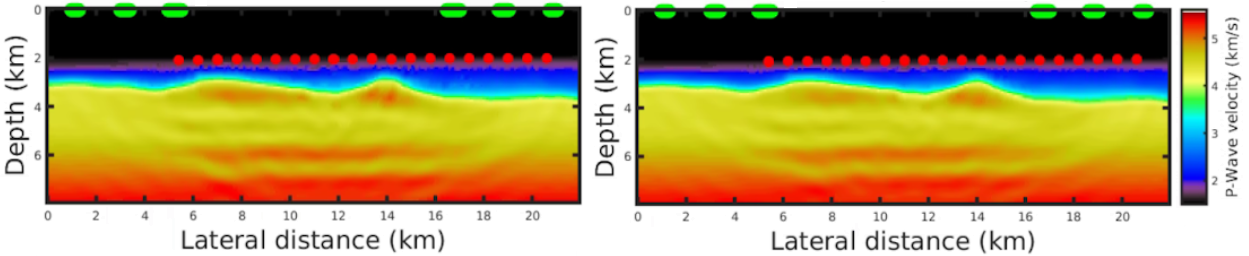}
\flushleft{\vspace{-.3cm}(h) \hspace{8cm} (i)}
\includegraphics[width=\textwidth]{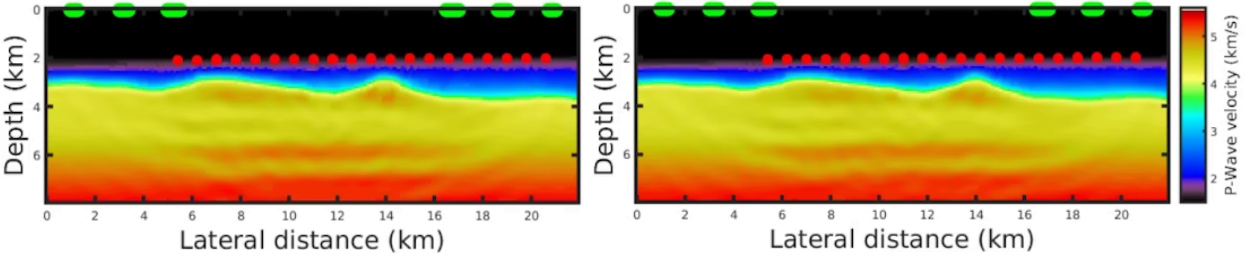}
\vspace{.05cm}
\caption{Vertical slices through the velocity model through the central crossline: (a) initial model; final FWI P-wave velocity models for the gradient without preconditioning case using (b) $L^1$ norm and (c) $L^2$ norm; the gradient preconditioned by source-receiver illumination using (d) $L^1$ norm and (e) $L^2$ norm; the gradient smoothed by applying the anisotropic nonstationary Bessel filter using (f) $L^1$ norm and (g) $L^2$ norm; and the gradient preconditioned by source-receiver illumination and anisotropic nonstationary Bessel filter using (h) $L^1$ norm and (i) $L^2$ norm. The low P-wave velocities recovered are seen within the pre-salt region, which is located at depths of more than $6 km$ and lateral extents of between $6$ and $16 km$. }
\label{fig:FWI_Results_CrossLine_NarrowTest}
\end{figure}

For simplicity, we shall restrict to showing vertical slices through the velocity models by the central crossline, where the initial model is depicted in Fig. \ref{fig:FWI_Results_CrossLine_NarrowTest}(a). Figures \ref{fig:FWI_Results_CrossLine_NarrowTest}(b) to \ref{fig:FWI_Results_CrossLine_NarrowTest}(i) show the FWI results for the four gradient preconditioning scenarios, in which the left and right sides refer to the $L^1$ and $L^2$ norm cases. The resulting models are significantly updated in the pre-salt region, as expected. The base of salt body is delineated, and the low P-wave velocity area is recovered. However, the reconstructed models obtained without the gradient preconditioning exhibit several artifacts especially close to the ocean floor (Figs. \ref{fig:FWI_Results_CrossLine_NarrowTest}(b) and \ref{fig:FWI_Results_CrossLine_NarrowTest}(c)). 

Indeed, the reconstructed models in the nonpreconditioned case show many artifacts independently of the objective function, being dominated by the imprint of the wavepaths, also introducing footprints in the pre-salt structures. In contrast, artifacts caused by wavepaths are attenuated when source-receiver illumination is considered (Figs. \ref{fig:FWI_Results_CrossLine_NarrowTest}(d) and \ref{fig:FWI_Results_CrossLine_NarrowTest}(e)). Only when the Bessel filter is used, such imprints are effectively mitigated (Figs. \ref{fig:FWI_Results_CrossLine_NarrowTest}(f) and \ref{fig:FWI_Results_CrossLine_NarrowTest}(g)). We also note that combining  the Bessel filter with the source-receiver illumination mainly mitigates the artifacts on top of salt (Figs. \ref{fig:FWI_Results_CrossLine_NarrowTest}(h) and \ref{fig:FWI_Results_CrossLine_NarrowTest}(i)). In this latter case, the reconstructed velocity model in the post-salt region is broadly similar to the initial model. This mitigation is due to the reduction of wavepath imprint after preconditioning the gradient by source-receiver illumination and gradient smoothing every FWI iteration and the data selection and acquisition geometry that prioritizes the deep penetrating diving waves.
\begin{figure}[!htb]
\centering
\begin{subfigure}[b]{0.4\textwidth}
\caption{}
 \includegraphics[width=\textwidth]{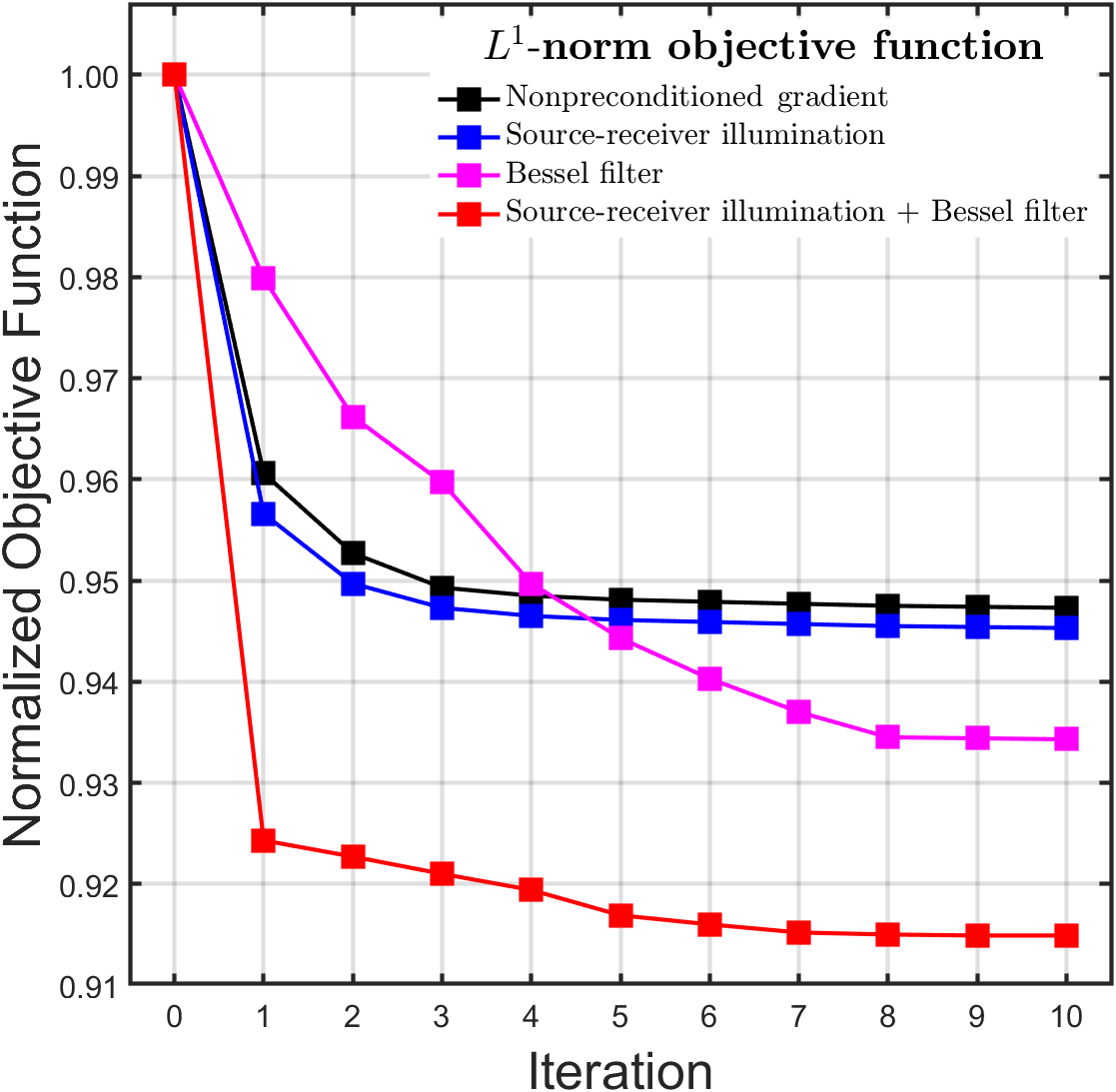}   
\end{subfigure}
\hfill
\begin{subfigure}[b]{0.4\textwidth}
\caption{}
 \includegraphics[width=\textwidth]{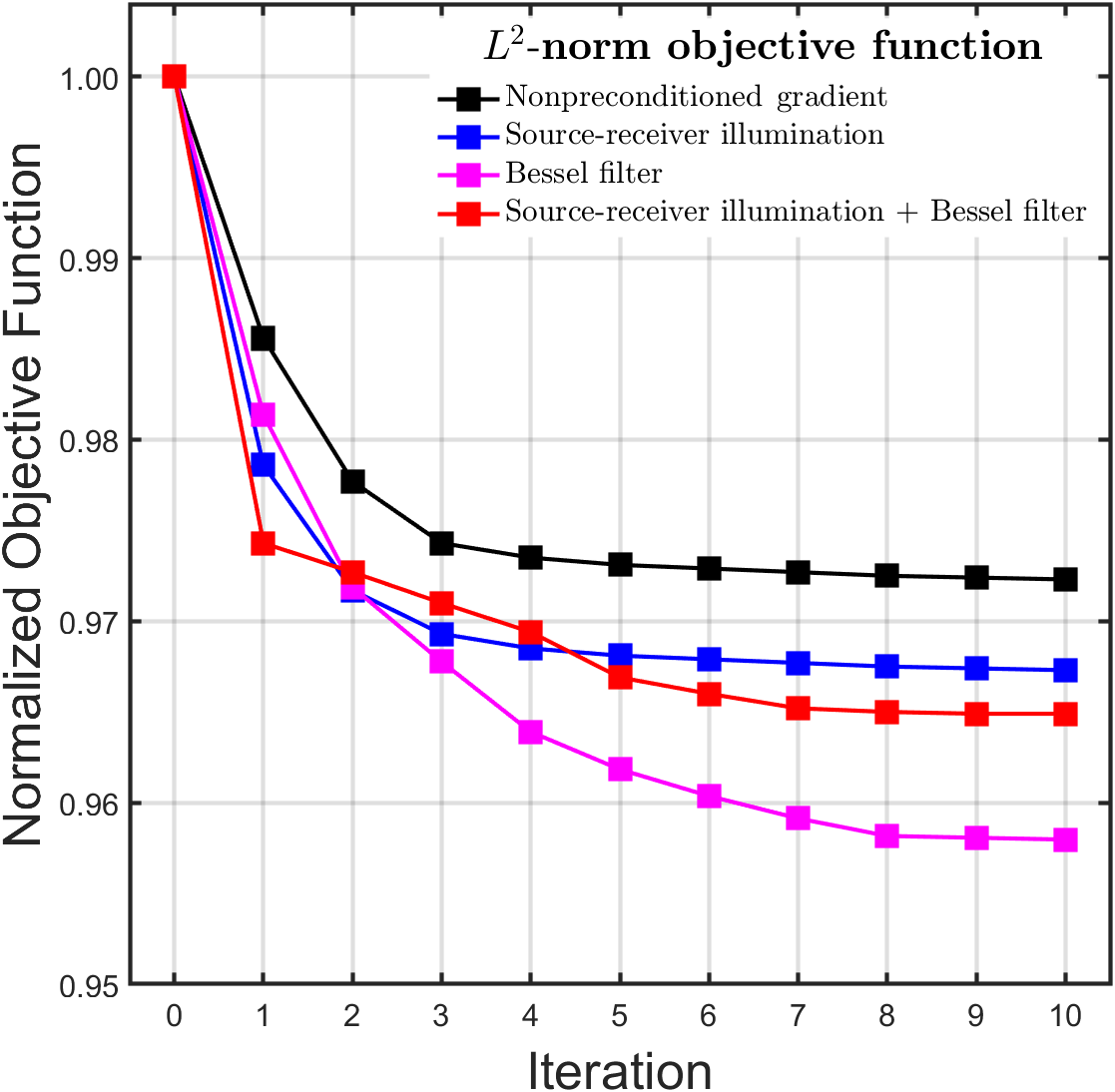}   
\end{subfigure}
\caption{FWI convergence curves of FWI around central crossline case study (narrow test) for the (a) $L^1$ and (b) $L^2$ norms.}
\label{fig:convergence_curve_narrow_test}
\end{figure}

Figure \ref{fig:convergence_curve_narrow_test} shows the FWI convergence curves for the four scenarios of gradient preconditioning, where panels (a) and (b) refer to the $L^1$ and $L^2$ norms, respectively. Since the objective functions based on the $L^1$ and $L^2$ norms have different sets of values, we normalize each objective function with respect to its first value. In all scenarios, the objective function based on the $L^1$ norm (Fig. \ref{fig:convergence_curve_narrow_test}(a)) exhibits a higher decrease rate than the objective function based on the $L^2$ norm (Fig. \ref{fig:convergence_curve_narrow_test}(b)). Moreover, it is worth emphasizing that the most significant decay is associated with the case of the FWI based on the $L^1$ norm with the gradient preconditioned by source-receiver illumination and anisotropic nonstationary Bessel filter, as depicted by the red curve in Fig. \ref{fig:convergence_curve_narrow_test}(a). Indeed, the more significant decay of the objective function based on the $L^1$ norm could mean that the FWI processing leads to a good retrieved model faster. This can be attributed to the low sensitivity of the $L^1$ norm to erratic data \cite{Brossier_et_al_2010_WhichResidualnorm,daSilva_et_al_2021_PSI_LaplaceDistribution}, which can handle the noise in the gradient calculation better than the $L^2$ norm.

\newpage
\subsubsection{FWI results assessment \label{sec:FWIResultsAssessment}}

In the circular shot OBN geometry, few common mid-points (CMPs) are available and are restricted to some areas of the studied region, making it suboptimal for migration purposes. Thus, to test the effectiveness of the FWI velocity models, we consider the conventional OBN dataset (with a 50m shot grid) that was acquired simultaneously with the circular shot geometry. Any improvements in the velocity models, while important on their own for reservoir characterization, should show corresponding improvements in the imaging of the pre-salt reservoirs. We use a reverse time migration (RTM) algorithm to compare the various FWI models in the seismic imaging domain. RTM algorithms are notable for their effectiveness in suppressing the effects of noise in the seismic image, velocity model inaccuracies, and for identifying complex structures such as the geometry of a salt body \cite{Zhou_et_al_2018_RTM_reviewZHOU2018207}.

We implemented an RTM algorithm using the inverse scattering imaging condition \cite{STOLK_et_al_2006_inversescattering} because the classical imaging condition may produce artifacts on a seismic image from undesired cross-correlations of, for instance, backscattered waves \cite{OPTROOT_et_all_20122_inverseScattering_RTM}. This imaging condition is based on the inverse scattering theory. It mitigates the backscattered waves by combining two separate seismic images, one of which is associated with the time derivatives of the downgoing source (incident wavefield) and the upgoing receiver fields (scattered wavefield), while the other concerning spatial gradients of the incident and scattered wavefields \cite{whitmore2012}.

\begin{figure}[!b]
    \centering \includegraphics[width=.35\columnwidth]{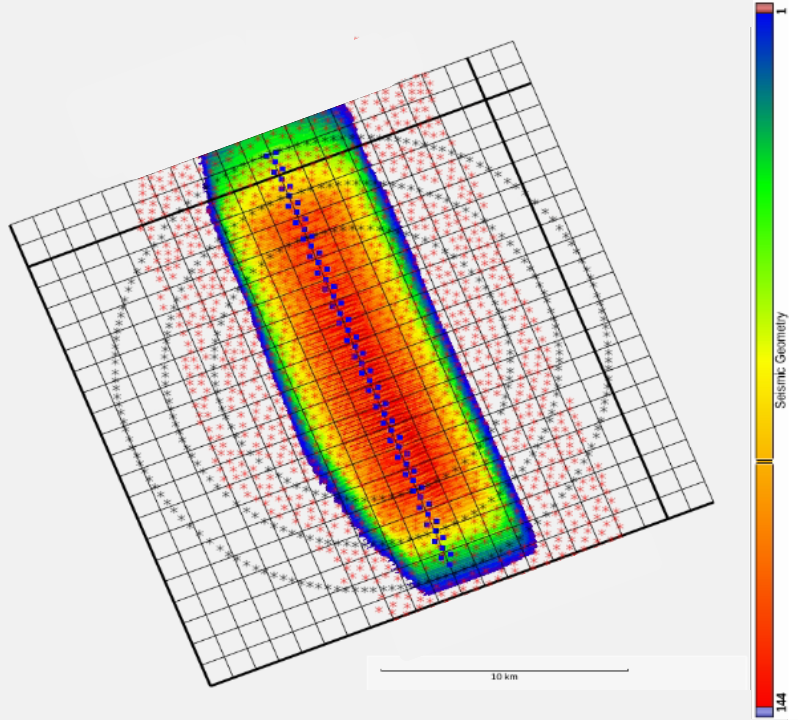}
    \caption{Fold map of the node lines used in the RTM imaging of the Brazilian pre-salt field. The colors vary from blue (low fold number) to red (high fold number). The blue dots are the node receiver positions, the red stars are the conventional OBN shot positions (that are used in the RTM image) and the black stars are the circular OBN shot positions (which are not used in this imaging exercise).}
    \label{fig:gdm_fold_map}
\end{figure}

Figure \ref{fig:gdm_fold_map} shows the shot-receiver distribution of the conventional OBN geometry, in which the blue dots and the red stars are the node and the conventional OBN source positions, respectively. In this figure, the colors vary from blue (low fold number) to red (high fold number), which shows the good fold of illumination along the node line in this "narrow azimuth test" example. In this study we only consider waveforms recorded up to a $7km$ maximum offset (source-receiver distance) to restrict the seismic data content to reflected waves, avoiding refracted waves that quickly reach the OBNs due to the high velocities present in the salt region.

\begin{figure}[!htb]
    \centering
    \includegraphics[width=.55\columnwidth]{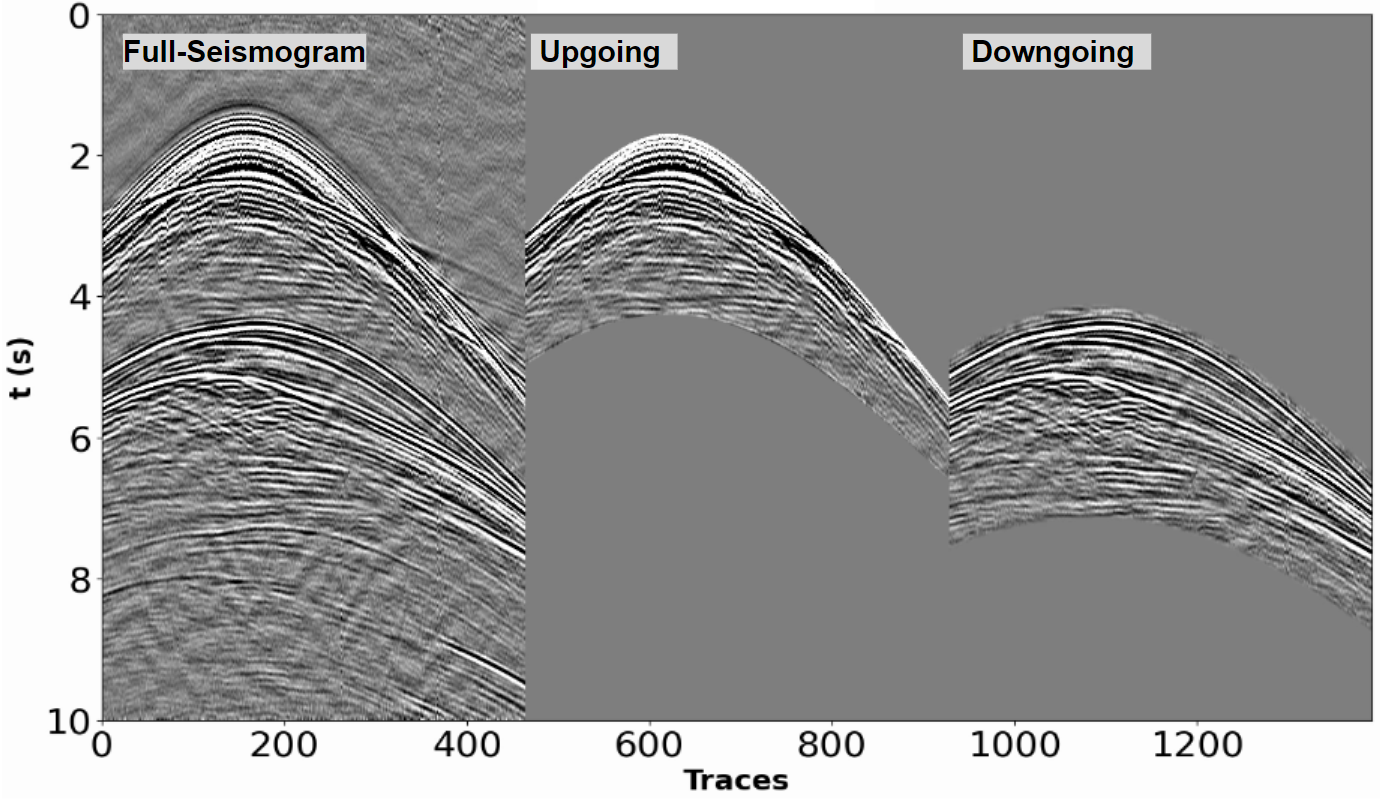}
    \caption{Seismograms with all events (left frame), upgoing shot-gather (center frame), and the downgoing gather (right frame). We use the downgoing and upgoing seismograms to generate the RTM images.}
    \label{fig:gdm_seismogram_division}
\end{figure}

Next we separate the shot gathers into upgoing and downgoing pressure wavefields. For this step, we compute traveltimes associated with the primary and the secondary ocean-floor multiple reflections to separate the recorded seismogram into upgoing wavefield (single reflection events) and downgoing wavefield (multiple reflection events). After separating the wavefields, we apply a low-pass filter ($0-15Hz$, Ormsby filter) on the dataset because simulating higher frequency content requires a denser finite-difference numerical grid, radically increasing the memory consumption and consequently making the 3D RTM running impractical with the available resources. Figure \ref{fig:gdm_seismogram_division} shows the filtered seismograms and the delimitation of the seismic data used in the migration process. The upgoing data is depicted in the center frame, while the downgoing content used for migration is in the right frame. We do not perform RTM using the second water bottom multiple due to the larger interference of elastic effects, which diminishes the amplitudes and pollutes the final image.

\begin{figure*}[!htb]
\flushleft{\hspace{-.1cm}(a) \hspace{8.4cm} (b)}
\includegraphics[width=\textwidth]{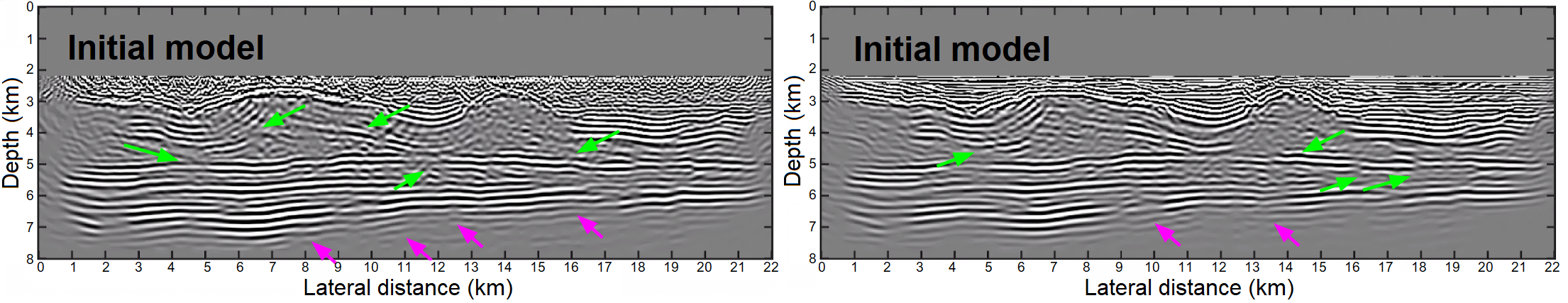}
\flushleft{\vspace{-.25cm}\hspace{-.1cm}(c) \hspace{8.4cm} (d)}
\includegraphics[width=\textwidth]{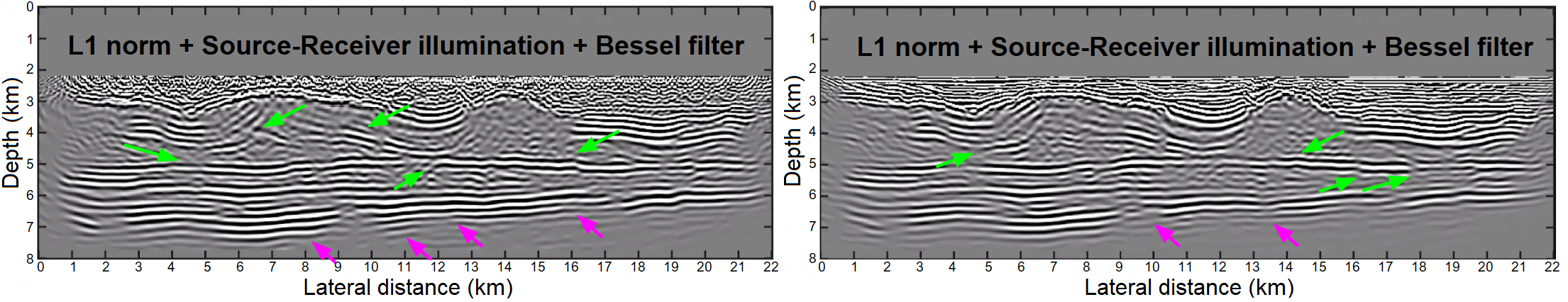}
\flushleft{\vspace{-.25cm}\hspace{-.1cm}(e) \hspace{8.4cm} (f)}
\includegraphics[width=\textwidth]{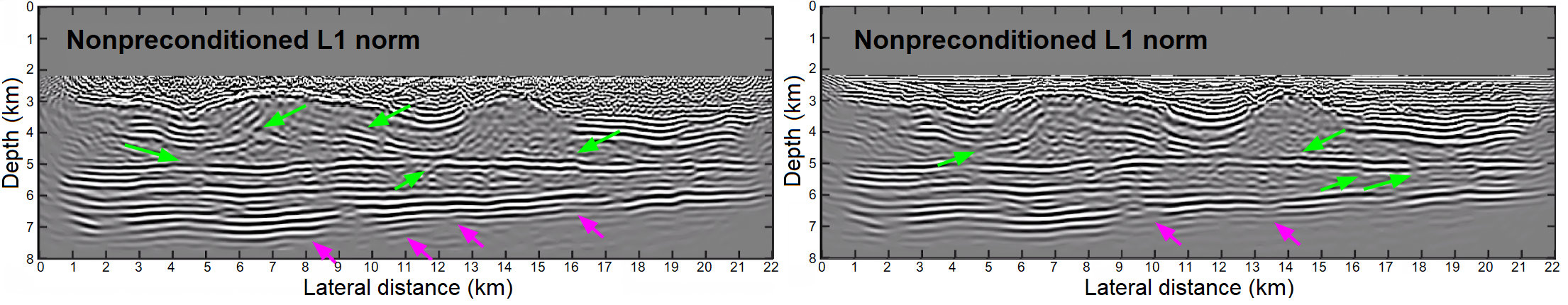}
\flushleft{\vspace{-.25cm}\hspace{-.1cm}(g) \hspace{8.4cm} (h)}
\includegraphics[width=\textwidth]{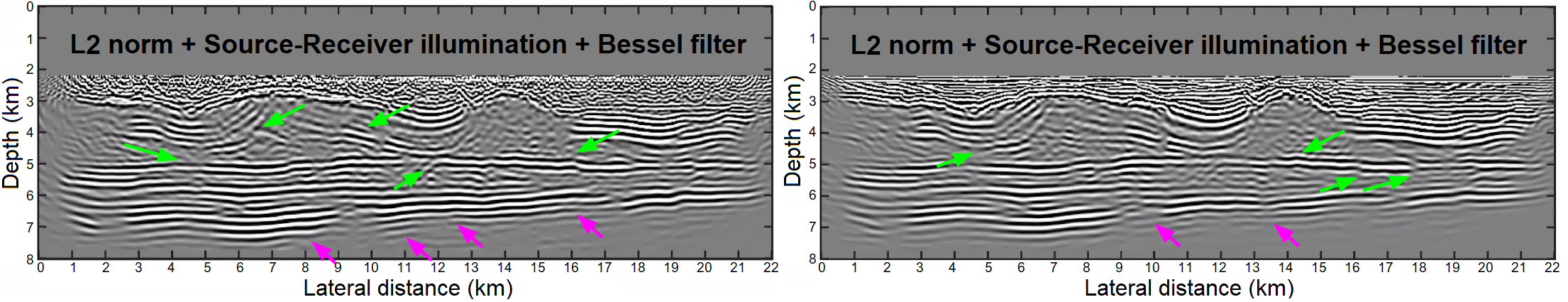}
\flushleft{\vspace{-.25cm}\hspace{-.1cm}(i) \hspace{8.5cm} (j)}
\includegraphics[width=\textwidth]{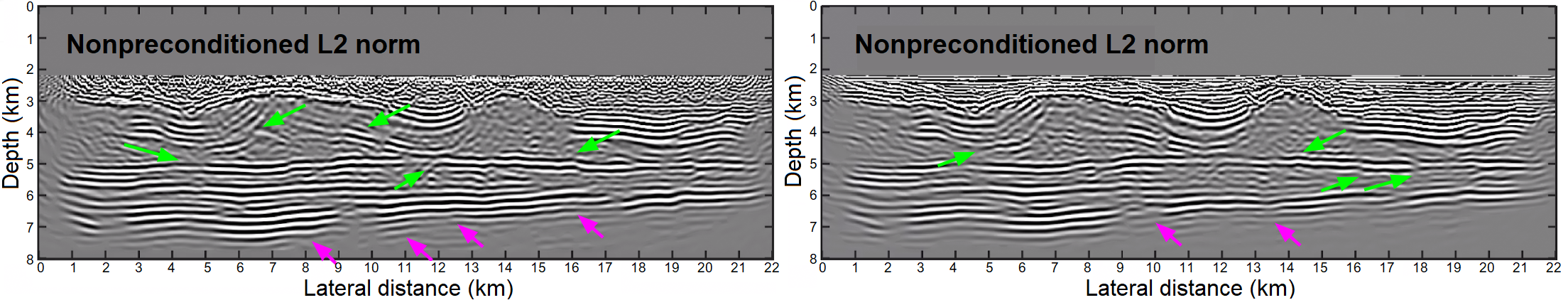}
\vspace{.025cm}
\caption{Traverses through the central crossline of the RTM images for the "narrow test" case. In the left column are the upgoing primary RTM images, and in the right column are the downgoing (mirror) RTM images for the (a)-(b) initial model; final FWI P-wave velocity models for the gradient preconditioned by source-receiver illumination and anisotropic nonstationary Bessel filter using (c)-(d) $L^1$ norm and (g)-(h) $L^2$ norm; The gradient without preconditioning case using (e)-(f) $L^1$ norm and (i)-(j) $L^2$ norm; The green arrows indicate the improvements on the salt body regarding the initial model, whilst the magenta ones refer to the improvements in the pre-salt layer continuity. \vspace{1cm}}
\label{fig:RTM_images_CrossLine_NarrowTest}
\end{figure*}

Considering the central crossline and its two neighboring crosslines from the conventional OBN acquisition, we obtained RTM images related to the initial model (Fig. \ref{fig:FWI_Results_CrossLine_NarrowTest}(a)) and the FWI resulting models (Fig. \ref{fig:FWI_Results_CrossLine_NarrowTest}(b)-(i)). We used an estimated seismic source following the methodology described in the previous section and a 3D time-domain acoustic modeling engine satisfying Eq. \eqref{eq:wave_equaqtion_time}. We resampled the P-wave velocity models to a regular grid spacing of $25m$. Figure \ref{fig:RTM_images_CrossLine_NarrowTest} shows the 15Hz RTM images, where the left column refers to the upgoing primary while the right column depicts downgoing (mirror) migrated images. Figures \ref{fig:RTM_images_CrossLine_NarrowTest}(a) and \ref{fig:RTM_images_CrossLine_NarrowTest}(b) show how robust the RTM algorithm is, as the pre-salt layers are highlighted even with an inaccurate model, which is the case of the initial FWI model. 

The annotated green arrows in Fig. \ref{fig:RTM_images_CrossLine_NarrowTest} indicate regions where the FWI models improve the body of salt of the initial model. Although our approach is target-oriented to the pre-salt region, it is possible to notice that the inverted models exhibit an improvement in the definition of stratification in the body of salt. Indeed, the RTM images with the FWI models highlight the salt stratification (Figs. \ref{fig:RTM_images_CrossLine_NarrowTest}(c)-(j)), which are significantly noisier in the initial model case (Figs. \ref{fig:RTM_images_CrossLine_NarrowTest}(a) and \ref{fig:RTM_images_CrossLine_NarrowTest}(b)). Also, the continuity of the salt stratification in the region with a $5km$ depth and $2km$ lateral distance is highlighted in the depth image depicted in Figs. \ref{fig:RTM_images_CrossLine_NarrowTest}(c), \ref{fig:RTM_images_CrossLine_NarrowTest}(e) and \ref{fig:RTM_images_CrossLine_NarrowTest}(g), are not present in the depth image with the initial model (Figs. \ref{fig:RTM_images_CrossLine_NarrowTest}(a) and \ref{fig:RTM_images_CrossLine_NarrowTest}(b)) and $L^2$ norm without preconditioning (Figs. \ref{fig:RTM_images_CrossLine_NarrowTest}(i) and \ref{fig:RTM_images_CrossLine_NarrowTest}(j)). Furthermore, the RTM images related to the FWI models highlight the salt base reflector, poorly represented in the initial model case, as indicated by the green arrows positioned on the referred layer at $\approx 5km$ depth. The annotated magenta arrows indicate regions where the FWI models improve the pre-salt area of the initial model. We notice that the inverted models exhibit improvements in the layer's continuity and show more pre-salt layers. In fact, the RTM images with the FWI models highlight the pre-salt layers (Figs. \ref{fig:RTM_images_CrossLine_NarrowTest}(c)-(j)), slightly noisier in the initial model case (Figs. \ref{fig:RTM_images_CrossLine_NarrowTest}(a) and \ref{fig:RTM_images_CrossLine_NarrowTest}(b)). 

\newpage
\newpage

\subsection{Circular shot OBN case study (full azimuth test)}

\begin{figure}[!htb]
\flushleft{(a) \hspace{8cm} (b)}
\includegraphics[width=\columnwidth]{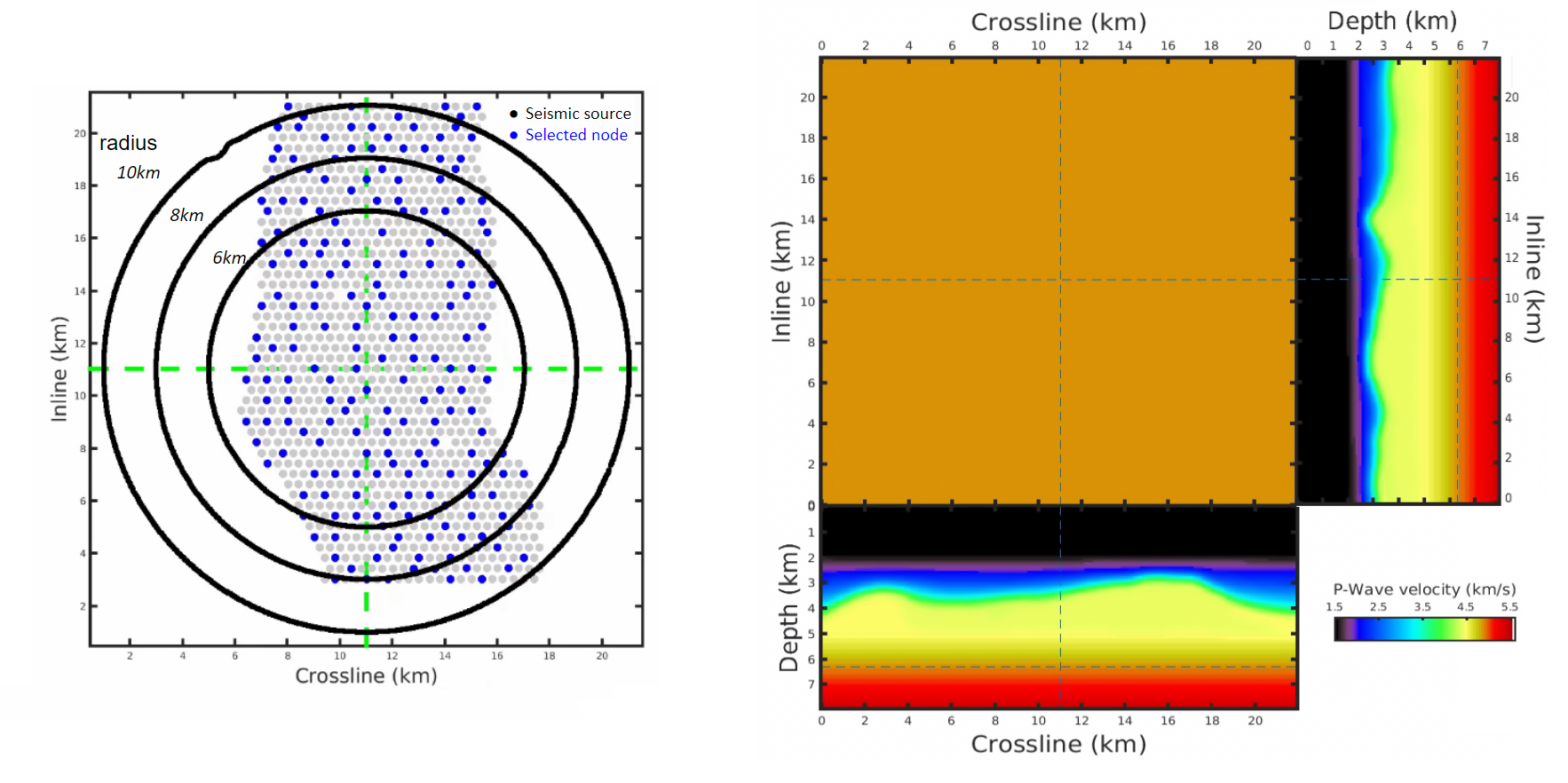}
\caption{(a) Base map of the subset of nodes used per FWI iteration, which comprises three concentric shooting circles (black curves) with radii of $6$, $8$ and $10km$, with a total of $3006$ seismic sources, regularly spaced every $50m$, and $200$ hydrophones (blue dots) over $44$ receiver lines. The dashed green lines indicate the central inline and crossline. (b) Depth slice through the pre-salt reservoir and vertical slices through the initial velocity model (Fig. \ref{fig:initial_model_acquistion_3Dfigure}) by the central inline and crossline.}
\label{fig:full_circular_shot_OBN_acquisition_selected_circularTEST}
\end{figure}

In this section we consider the full-azimuth coverage provided by the circular shot OBN dataset. In this study we use a subset of nodes per FWI iteration, chosen randomly, since using all nodes in each FWI iteration is inefficient in the sense of computational processing effort \cite{Warner_et_al_2013_Geophysics_AnisotropicFWI,Kamath_et_al_2021_Geophysics_FWI_Multiparameter} and may produce regular interference patterns \cite{Diaz_Guitton_2011_SEG_RandomShotDecimation}. For each FWI iteration we select $200$ nodes randomly using a 2D uniform distribution by imposing that each receiver covers a circular area with a radius of $450m$ (avoiding neighbor receivers in the same FWI iteration), as depicted in Fig. \ref{fig:full_circular_shot_OBN_acquisition_selected_circularTEST}(a).

Next we perform 3D time-domain FWI by employing objective functions based on $L^1$ and $L^2$ criteria, in which we set up $10$ iterations for all data inversions. We also consider two gradient preconditioning scenarios: (i) a gradient without preconditioning; and (ii) a gradient preconditioned by source-receiver illumination and smoothed through the anisotropic nonstationary Bessel filter, as described in the previous sections. Regardless of the gradient preconditioning scenario, each FWI simulation takes about 27 hours. Figure \ref{fig:full_circular_shot_OBN_acquisition_selected_circularTEST}(b) shows a schematic map illustrating the initial model.

\begin{figure*}[!t]
     \flushleft{(a) \hspace{8.4cm} (b)}
     \includegraphics[width=\textwidth]{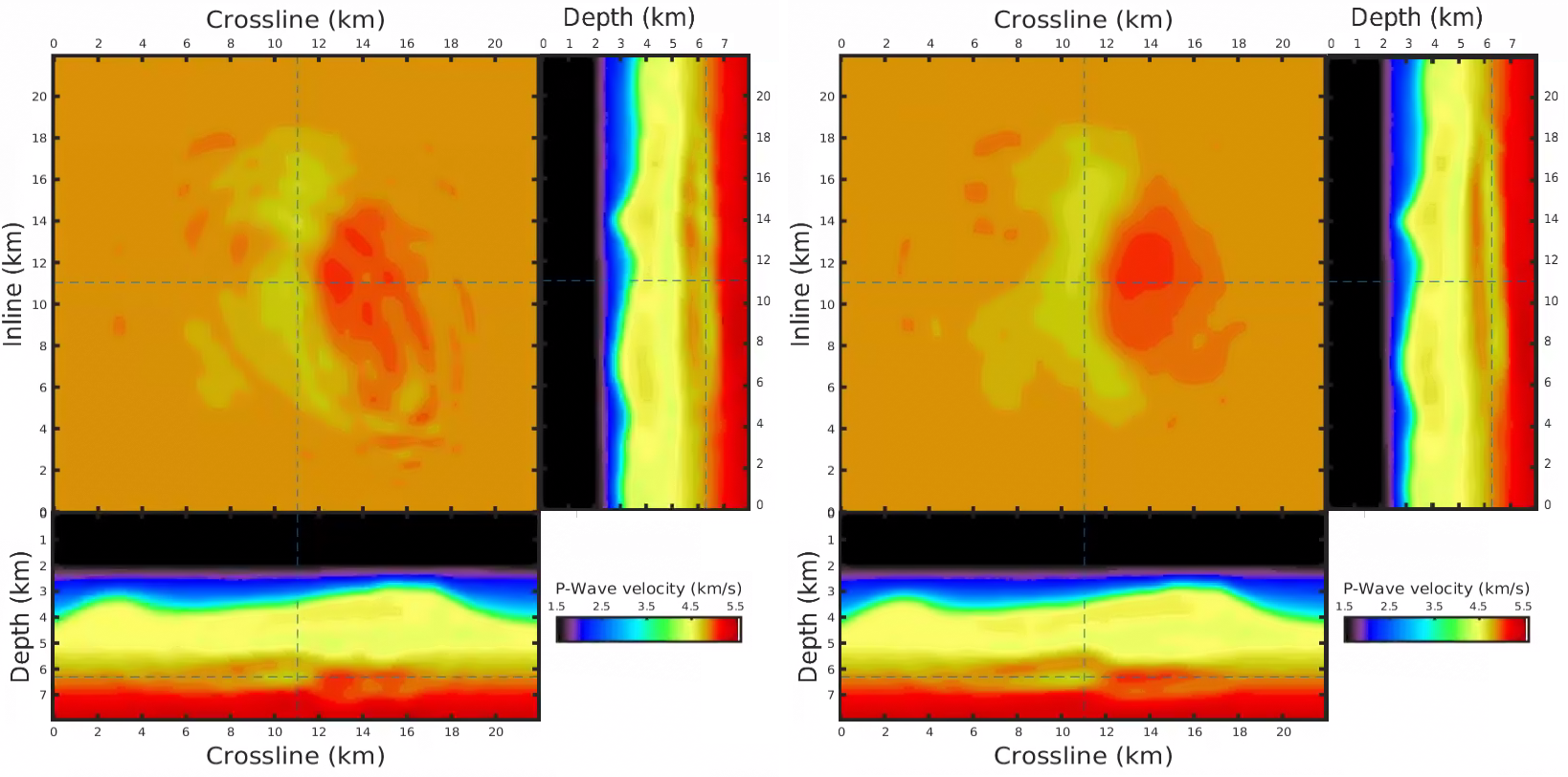}
     \flushleft{\vspace{-.25cm}(c) \hspace{8.4cm} (d)}
     \includegraphics[width=\textwidth]{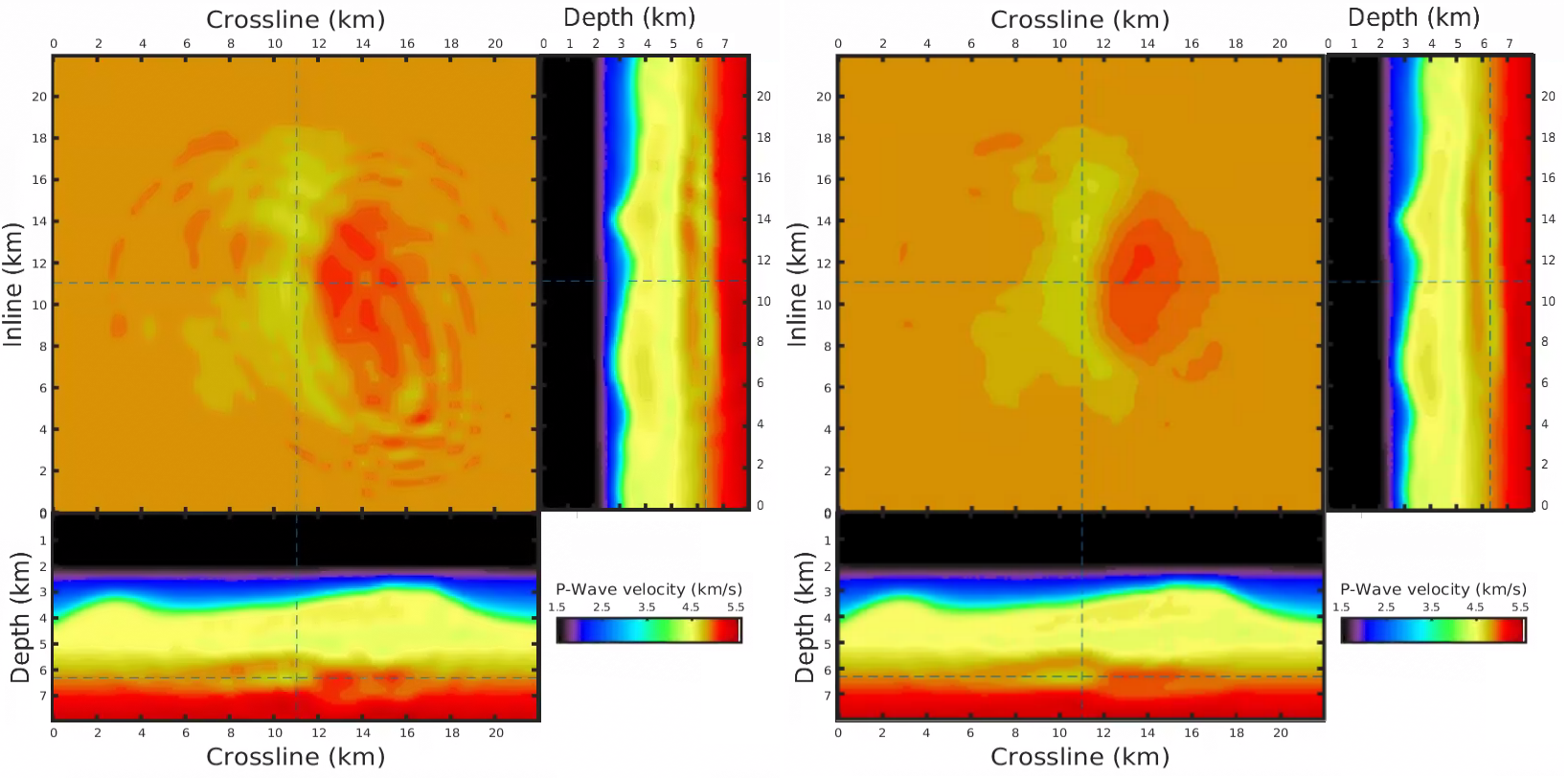}
        \caption{Depth slice through the pre-salt reservoir and vertical slices through the final FWI velocity models by the central inline and  crossline for the preconditioned gradient case using the (a) $L^2$ norm and (b) $L^1$ norm, and the nonpreconditioned gradient case using (c) $L^2$ norm and (d) $L^1$ norm. The low P-wave velocities recovered are seen within the pre-salt region.}
        \label{fig:circular_FWI_results}
\end{figure*}

Figure \ref{fig:circular_FWI_results} shows the FWI results, in which we depict a depth slice through the pre-salt reservoir and vertical slices through the central inline and crossline. The left and right columns refer to the $L^2$ and $L^1$ norms cases, respectively. As in the "narrow test" study presented above, the resulting models are significantly updated in the pre-salt region by delineating the base of salt body and the low P-wave velocity deep region. However, regardless of whether the gradient was preconditioned, we notice that the P-wave models produced by the FWI based on the $L^2$ norm exhibit several artifacts when taking into account the full azimuthal coverage, as depicted in Figs. \ref{fig:circular_FWI_results}(a) and \ref{fig:circular_FWI_results}(c). In fact, the FWI models reconstructed using the $L^2$ norm present the imprint of the wavepaths in the pre-salt region. This is because of the weights on the residual data (the difference between the modeled and observed data, also known as errors) given by the objective function based on the $L^2$ norm, as discussed by Ref.   \cite{dosSantosLima_et_al_2023_PlosOne}. In this regard, analyzing Eq. \eqref{eq:adjoint_wave_equation_time} for $p=2$, it is noticeable that the $L^2$ objective function suppresses small errors and magnifies large errors in the inversion process since the associated adjoint source is proportional to the residual data.
 \clearpage

On the other hand, we notice that the strong wavepath footprints observed at the top of salt in the previous case study (Fig. \ref{fig:FWI_Results_CrossLine_NarrowTest}) were mitigated regardless of the objective function employed, which suggests that this was a beneficial impact of the azimuthal coverage of the circular shot OBN geometry. Regarding the FWI results based on the $L^1$ norm (Figs. \ref{fig:circular_FWI_results}(b) and \ref{fig:circular_FWI_results}(d)), we notice that they exhibit less artifacts than the $L^2$ norm (Figs. \ref{fig:circular_FWI_results}(a) and \ref{fig:circular_FWI_results}(c)), by effectively alleviating the wavepath imprints and giving a broader and sharper representation of the pre-salt region, especially when the preconditioned gradient is employed (Figs. \ref{fig:circular_FWI_results}(b)). This alleviating is because the $L^1$ objective function handles the noise better in the gradient computation than the $L^2$ norm \cite{Brossier_et_al_2010_WhichResidualnorm,daSilva_et_al_2021_PSI_LaplaceDistribution}.

\begin{figure}[!b]
     \includegraphics[width=\columnwidth]{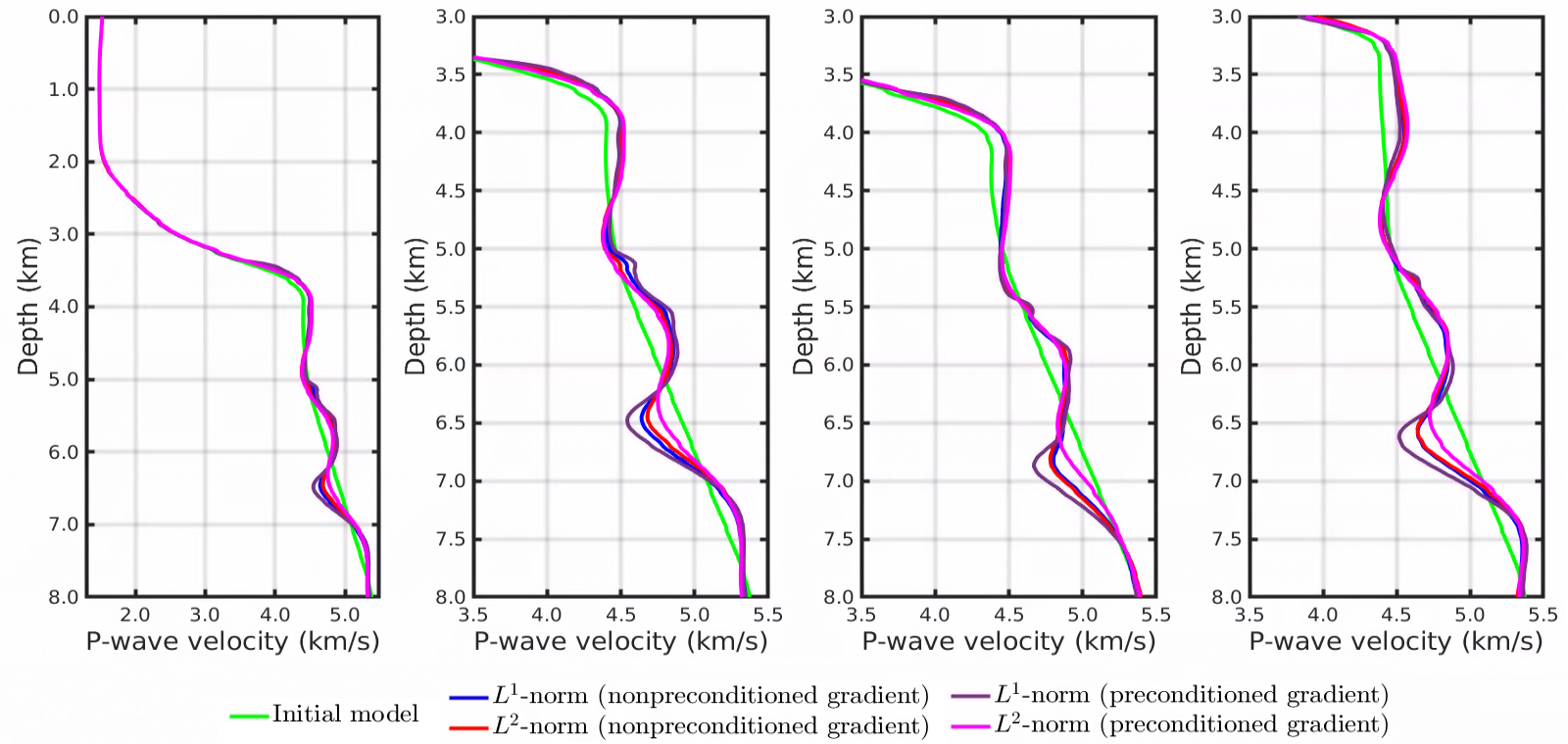}
     \caption{ Vertical profiles associated with initial velocity models (green curve) and FWI models. Panels (a) and (b) refer to a profile located in the center of the shot circles, i.e., (crossline, inline) = (11, 11) km. Panel (b) is a zoom of the profile illustrated in Panel (a). Panels (c) and (d) are velocity profiles located at the positions (crossline, inline) = (8, 11) km and (crossline, inline) = (11, 8) km.}
        \label{fig:circular_FWI_results__verticalprofiles}
\end{figure}

Figure \ref{fig:circular_FWI_results__verticalprofiles} shows the vertical profiles of the initial and FWI models, with panel (a) representing the profile at the center of the circles. We notice that up to just over 3 km of depth, all the models are similar, as depicted in Fig. \ref{fig:circular_FWI_results__verticalprofiles}(a). This is because we analyzed only the diving waves, and therefore, the best-illuminated seismic regions in this case are the deeper regions. Figure \ref{fig:circular_FWI_results__verticalprofiles}(b) shows a zoom of Fig. \ref{fig:circular_FWI_results__verticalprofiles}(a), where it is possible to see that all approaches update the velocities at the top of the salt (around 4 km depth). In addition, all approaches update the low-velocity region in the pre-salt area (around 6.5 km depth), demonstrating a consistent trend in updating the models via FWI. However, it is worth highlighting that the objective function based on the $L^1$ norm, combined with gradient preconditioning, was able to distinguish this low-velocity region more clearly than the other approaches, as depicted by the purple line in Fig. \ref{fig:circular_FWI_results__verticalprofiles}. It is also worth noting that the other vertical profiles support these results, as shown in Figs. \ref{fig:circular_FWI_results__verticalprofiles}(c) and \ref{fig:circular_FWI_results__verticalprofiles}(d).

\begin{figure}[!htb]
\centering
     \includegraphics[width=.5\columnwidth]{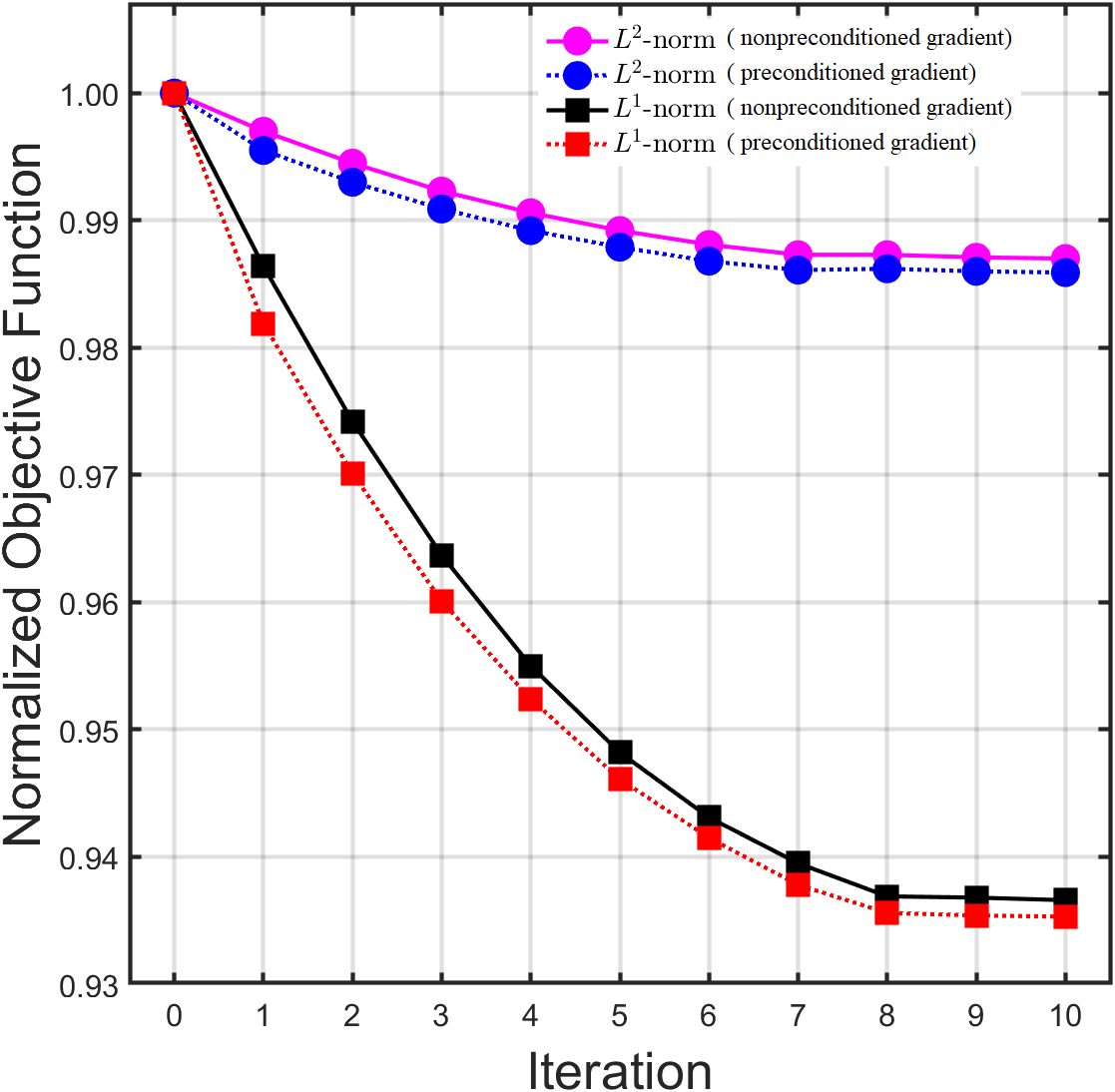}
     \caption{FWI convergence curves of the circular shot OBN case study (full azimuth test).}
        \label{fig:circular_FWI_results__convergence_curve}
\end{figure}

Figure \ref{fig:circular_FWI_results__convergence_curve} shows the FWI convergence curves. The objective function based on the $L^1$ norm (represented by the black and red squares in Fig. \ref{fig:circular_FWI_results__convergence_curve}) decays faster than the objective function based on the $L^2$ norm (represented by the magenta and blue circles in Fig. \ref{fig:circular_FWI_results__convergence_curve}). In addition, it is worth noting that applying preconditioning to the gradient leads to a slightly faster decay of the objective function than in the scenario where no preconditioning is considered.

\begin{figure}[!htb]
     \flushleft{\hspace{3.7cm} (a) \hspace{3.7cm} (b)\\}
     \centering
     \includegraphics[width=.5\columnwidth]{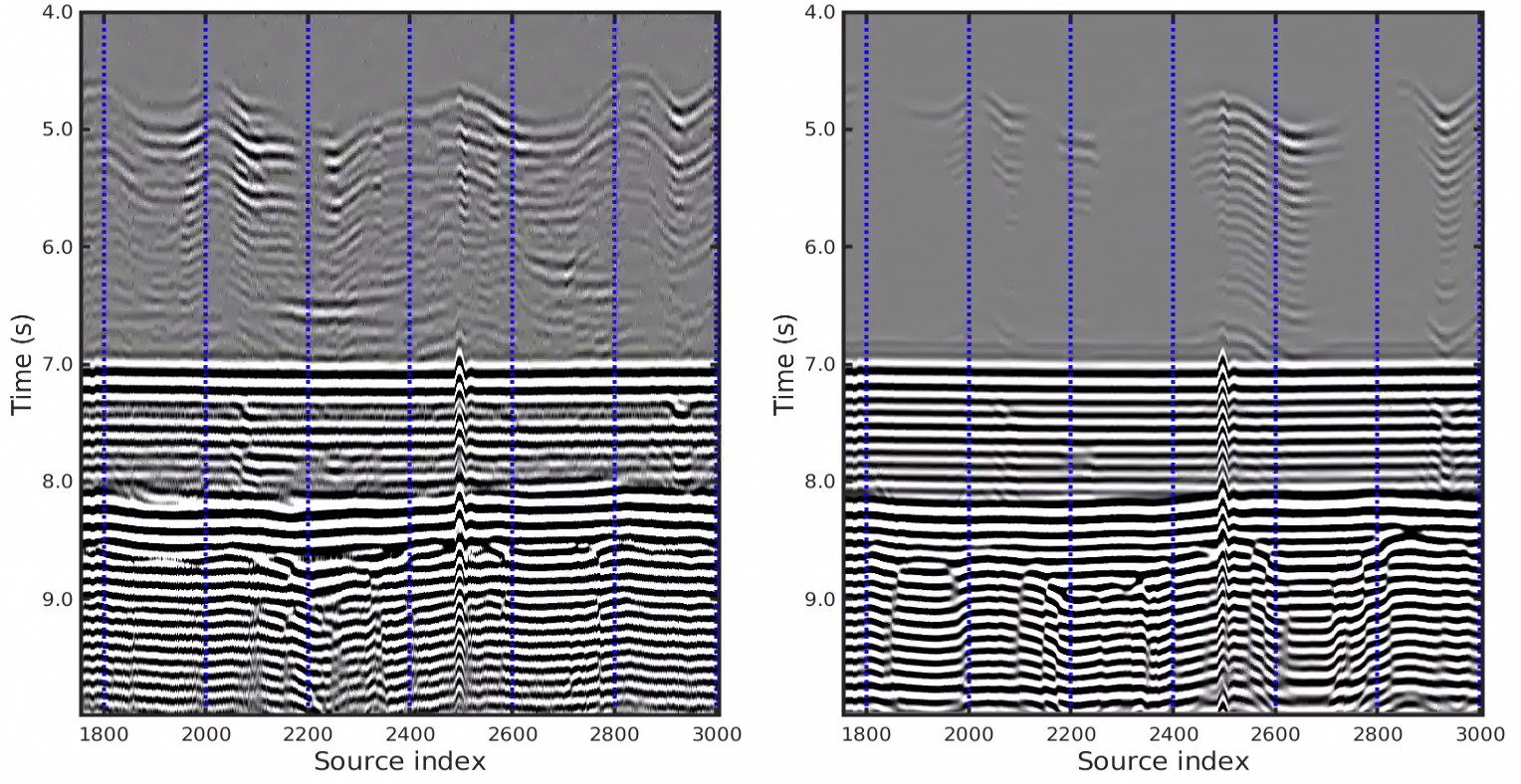}
     \flushleft{\vspace{-.25cm}\hspace{3.7cm}(c) \hspace{3.7cm} (d)\\}
     \centering
     \includegraphics[width=.5\columnwidth]{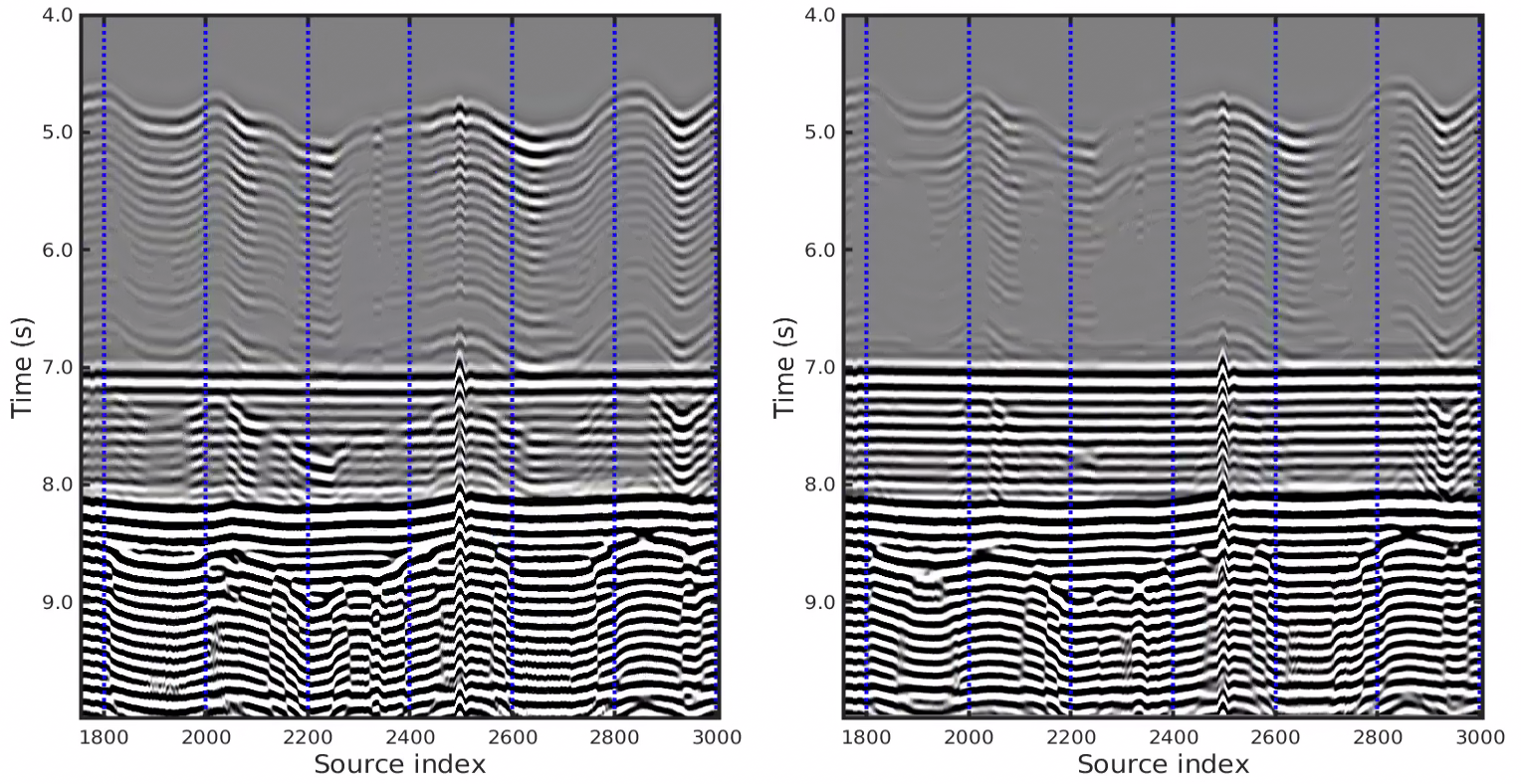}
     \flushleft{\vspace{-.25cm}\hspace{3.7cm}(e) \hspace{3.7cm} (f)\\}
     \centering
     \includegraphics[width=.5\columnwidth]{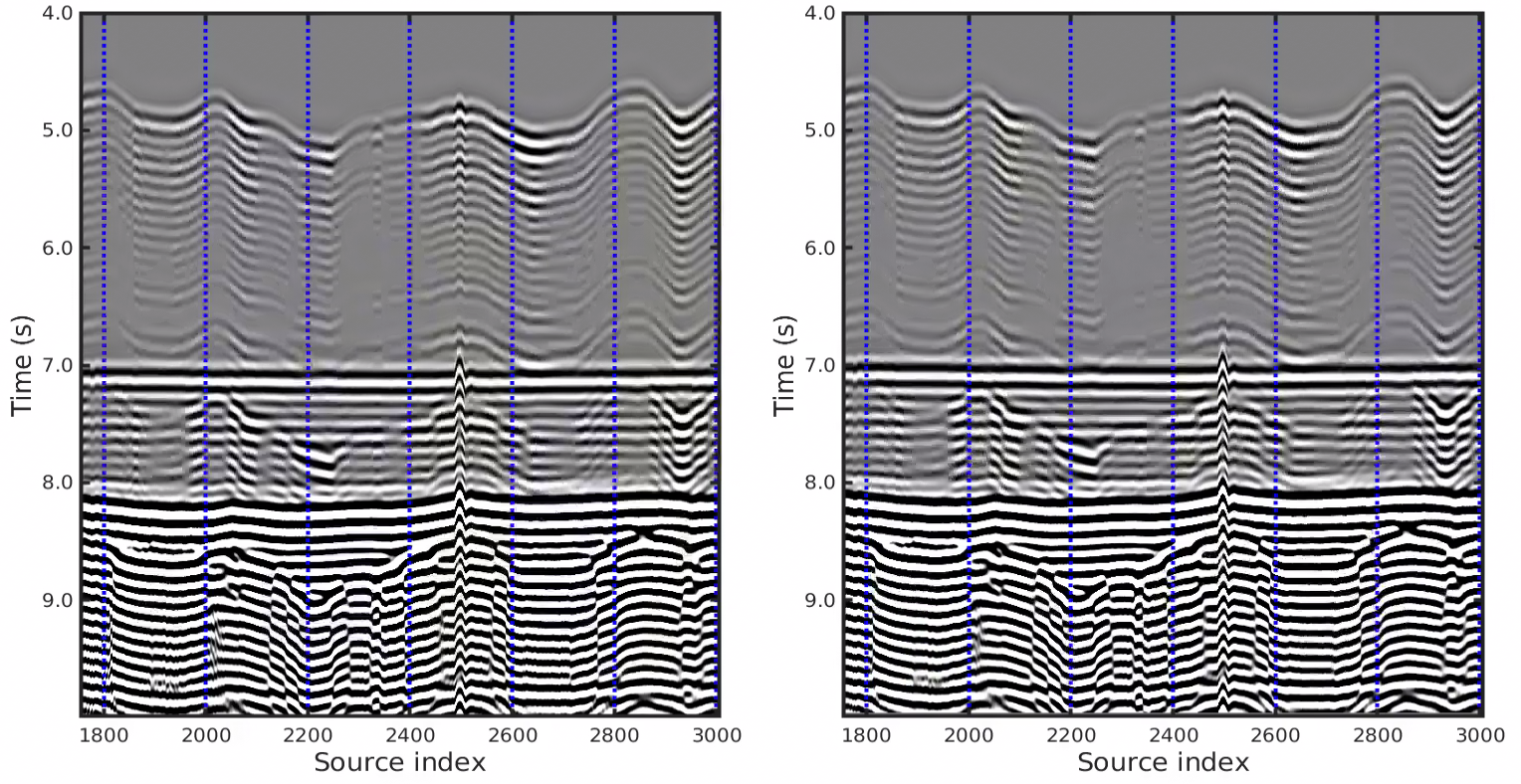}
        \caption{Receiver gathers corresponding to the node at the center of the shot concentric circles (central node) and the seismic sources of the larger-radius circle (10 km), in which panel (a) refer to the observed data and (b) to the modeled data using the initial model.  Panels (c) and (d) show the modeled data using the reconstructed models with the $L^1$ norm, without and with gradient preconditioning, respectively. Panels (e) and (f) show the modeled data using the reconstructed models with the $L^2$ norm, without and with gradient preconditioning, respectively. }
        \label{fig:circular_FWI_seismograms_circle10_centralnode}
\end{figure}

Figure \ref{fig:circular_FWI_seismograms_circle10_centralnode} shows the receiver gathers corresponding to the node at the center of the shot concentric circles (central node) and the seismic sources of the larger-radius circle (10 km). Panels (a) and (b) display the observed data and the modeled data using the initial model, respectively. Panels (c) and (d) show the modeled data using the reconstructed models with the $L^1$ norm, without and with gradient preconditioning, respectively. Panels (e) and (f) present the modeled data using the reconstructed models with the $L^2$ norm, without and with gradient preconditioning, respectively. In this figure, to provide a clearer visualization of the first arrivals, we adjusted the scales of all receiver-gather images so that the minimum and maximum values displayed correspond to the 3rd and 97th percentiles of the data. This way, we highlight the most relevant waveforms considered in our diving wave analysis, especially the arrivals before the direct wave ($\approx 7 s$). We notice that the retrieved models improve the capture of the main characteristics of the observed data (Fig. \ref{fig:circular_FWI_seismograms_circle10_centralnode}(a)), correcting the initial model's shortcomings. Indeed, in all inversion cases, the resulting receiver-gathers closely resemble the observed data when compared with the initial model case (Fig. \ref{fig:circular_FWI_seismograms_circle10_centralnode}(b)). For instance, FWI analysis satisfactorily filled the waveform gap between 4 and 6 s present in the initial model case, yielding receiver-gathers that are more consistent with observations. However, we highlight that the $L^1$ norm case with a preconditioned gradient (Fig. \ref{fig:circular_FWI_seismograms_circle10_centralnode}(d)) shows a greater similarity to the observations (Fig. \ref{fig:circular_FWI_seismograms_circle10_centralnode}(a)). For example, the amplitude variations of the first arrival, associated with source indices between 2000 and 2200, demonstrate this. Furthermore, note that some strongly visible features in the observed data (Fig. \ref{fig:circular_FWI_seismograms_circle10_centralnode}(a)) are not apparent in the modeled data. For instance, the most notable event between times 6 and 7 at source index 2200 is not easily discernible in the modeled receiver gathers (Figs. \ref{fig:circular_FWI_seismograms_circle10_centralnode}(c)-(f)). This is because more sophisticated modeling mechanisms need to be considered in the next steps of inversion at higher frequencies, as high energy is associated with density and elastic effects, as discussed by Ref. \cite{daSilva_et_al_2022_EAGE_KleinGordon}.

In the narrow test example the primary-RTM image was able to give a wider image of the pre-salt region than the mirror-RTM image. Thus, we present here only the upgoing primary-related depth image to summarize our results. We use the same datasets presented in section \ref{sec:FWIResultsAssessment}  to run 3D RTM applications associated with the conventional OBN acquisition. In this regard, we obtained upgoing primary RTM images associated with the initial model and the FWI resultant models by considering the three crosslines from the conventional OBN geometry.

\clearpage

\begin{figure*}[!htb]
\flushleft{\hspace{3.7cm}(a)}
\includegraphics[width=\textwidth]{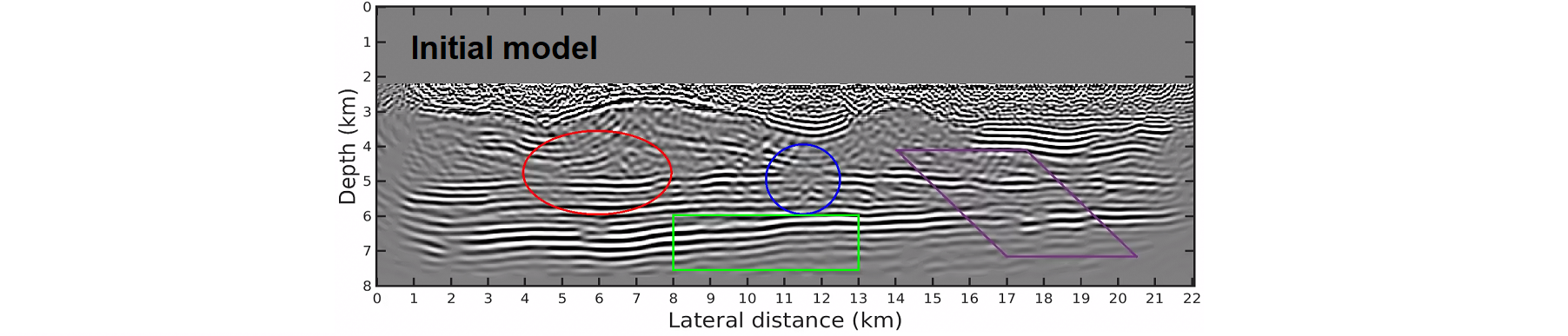}
\flushleft{\vspace{-.2cm}(b) \hspace{8.3cm} (c)}
\includegraphics[width=\textwidth]{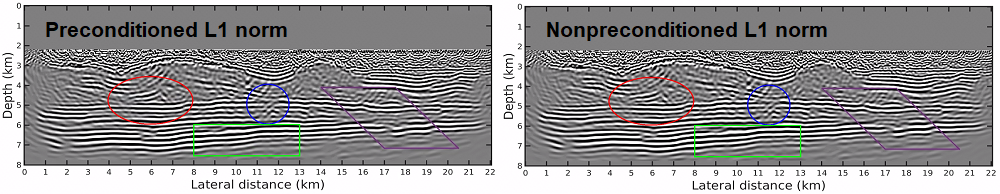}
\flushleft{\vspace{-.2cm}(d) \hspace{8.3cm} (e)}
\includegraphics[width=\textwidth]{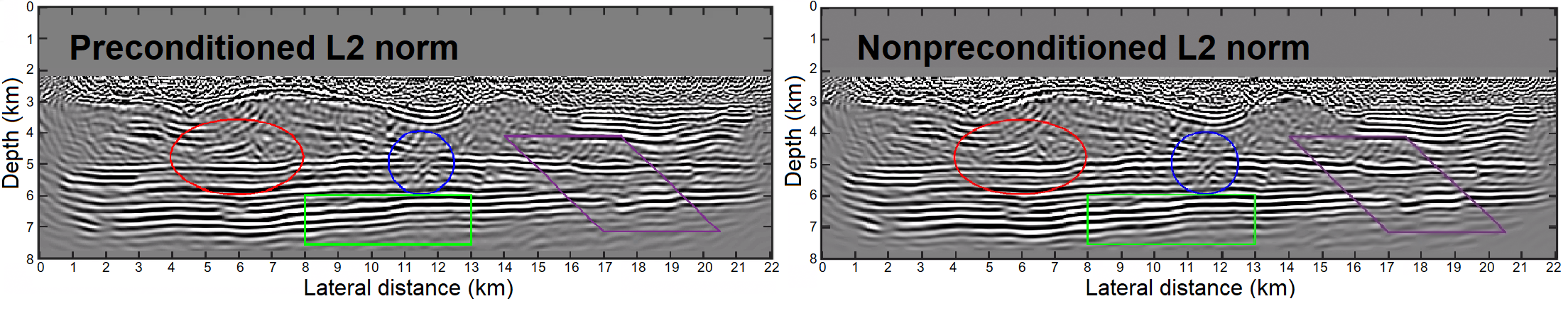}
\caption{Vertical slices through the upgoing RTM images through the central crossline for the (a) initial model; final FWI P-wave velocity models for the gradient preconditioned by source-receiver illumination and anisotropic nonstationary Bessel filter using (b) $L^1$ norm and (d) $L^2$ norm; The gradient without preconditioning case using (c) $L^1$ norm and (e) $L^2$ norm.}
\label{fig:RTM_images_CrossLine_CircularTest}
\end{figure*}

Figure \ref{fig:RTM_images_CrossLine_CircularTest} shows 15 Hz RTM images using the initial model and the FWI resulting models. Indeed, the annotated geometric shapes show discontinuities or the absence of layers in the RTM image associated with the initial model. In contrast, the base of salt and pre-salt regions of the RTM stack exhibit the most evident model updates after the application of FWI (blue circles and green squared), in which the base of salt is better identified (the resolution of the pre-salt layers is improved) and the reflectors are seen more continuous and focused. Furthermore, the salt stratification is better defined and less noisy in the FWI results (red ellipse). The results shown in Fig. \ref{fig:RTM_images_CrossLine_CircularTest} are the first to demonstrate the value of the circular OBN geometry to improve reservoir imaging, in particular to increase the lateral resolution in the pre-salt region.

\clearpage

\section{Final remarks} \label{sec:finalremarks}

In this work we have evaluated the capability of the 3D acoustic time-domain FWI to estimate P-wave velocities in the Brazilian pre-salt oil region using the recently introduced circular shot OBN acquisition geometry. In this regard, we introduced a modeling and inversion workflow to deal with non-preprocessed OBN refraction data from hydrophone components using a single GPU as the main computational resource. Our proposed FWI workflow consists of: (i) an automatic data selection using the Eikonal equation; (ii) data filtering using an Ormsby bandpass filter; (iii) source estimation considering the filtered field data and the acoustic wave propagation modeling operator; (iv) data resampling accordingly to the Nyquist–Shannon sampling; (v) random shot sampling; (vi) source-receiver reciprocity; (vii) nonlinear optimization of objective functions; and (viii) a gradient conditioned by the source-receiver illumination and anisotropic nonstationary Bessel filter. 

The Brazilian pre-salt case study revealed that the circular shot OBN geometry was fundamental for providing the mostly refracted waves separately from reflected waves due to its full azimuthal coverage and ultra-long offsets. This acquisition geometry allowed performing the low-frequency FWI successfully. In this regard, waveforms and velocity model resampling, along with the gradient calculation using random boundaries, allowed to use of the memory available in a single GPU, reducing computational costs, a significant problem in FWI issues. Indeed, the low-cost circular geometry proved to be efficient for illuminating the pre-salt region via deep penetrating diving waves, providing waveforms recorded in the far field in the FWI process, as predicted in previous synthetic studies \cite{Costa_et_al_2020_SEG_wavepaths,Lopez_et_al_2020_SEG_OBNcircular,daSilva_et_al_SEG_2021_qFWI_BrazilianPreSalt}.  We further showed that combining source-receiver illumination energy and the anisotropic nonstationary Bessel filter, in the gradient preconditioning, provides significant mitigation of the acquisition footprint without degrading the deeper structures. Indeed, the resulting models show that our proposed  workflow is an effective strategy to deal with wide-aperture acquisition geometries such as the circular shot OBN.

Furthermore, in the inversion process, we employed the two most popular objective functions in the literature based on the $L^1$ and $L^2$ norms. The RTM images showed that the reflectivity models generated by both approaches are similar. However, the objective function based on the $L^1$ norm generates P-wave velocity models with fewer artifacts than the $L^2$ norm case, especially when full azimuthal coverage is considered. 

Despite the recognition of the importance of long-offset illumination in enhancing the quality of velocity models \cite{Roende_et_al_2023_OBNacquis}, the acquisition of such surveys may be inefficient, requiring multiple source vessels to reach the very large offsets of interest (as far as 40 km or more). Using instead a circular geometry, only the long offsets of interest are acquired, which are used to supplement the FWI velocity models obtained from the conventional OBN shooting geometry. This supplementary long-offset illumination can be acquired in a relatively short time and should be considered in every OBN seismic campaign.

As perspectives, we intend to investigate the applicability of our proposal to carry out inversions with higher frequency content (following the multiscale FWI workflow), in which viscoelastic effects have a strong influence on the time and amplitudes of the waveforms, by combining more sophisticated objective functions \cite{Barnier_et_al_2023_modelextension,Jiang_et_al_2023_IEEE} with robust target-oriented approaches \cite{Barbosa_et_al_2022_TargetOrientedFWIGreen,Biondi_et_al_2023_TargetOrientedElastic}.






\section*{Acknowledgments}
The authors from Fluminense Federal University (UFF) gratefully acknowledge the support from Shell Brasil Petróleo Ltda through the R\&D project “Refraction seismic for pre-salt reservoirs” (ANP no. 21727-3). The strategic importance of the R\&D levy regulation from the National Agency for Petroleum, Natural Gas and Biofuels (ANP) is also appreciated. We would like to thank Rodrigo S. Stern for the crucial IT support.

\end{document}